# Conformationally frustrated epitaxy at heteromolecular interface

*Jakub Planer,[1] Pavel Procházka,[1] Tomáš Krajňák,[1] Veronika Stará,[1] Daniela Hrubá, [1] Zdeněk Endstrasser, [1] Tomáš Skála,[2] Jan Čechal[1,3]**

[1] CEITEC - Central European Institute of Technology, Brno University of Technology, Purkyňova 123, 612 00 Brno, Czech Republic.

[2] Department of Surface and Plasma Science, Faculty of Mathematics and Physics, Charles University, V Holešovičkách 2, 180 00 Prague 8, Czech Republic.

[3] Institute of Physical Engineering, Brno University of Technology, Technická 2896/2, 616 69 Brno, Czech Republic.

**Abstract**

Precise control of molecular heteroepitaxy offers a route to multilayer architectures with well-defined interfaces. However, the role of intramolecular flexibility in epitaxial matching remains insufficiently understood. Here, we investigate the heteroepitaxy of intact 1,3,5-tris(4-carboxyphenyl)benzene (BTB) on a molecular δ-layer composed of fully deprotonated BTB molecules on Ag(111). Combining scanning tunneling microscopy, low-energy electron microscopy and diffraction, X-ray photoelectron spectroscopy, and density functional theory, we identify the key interactions controlling molecular packing and epitaxial growth. We show that intermolecular bonding motifs are preserved on the molecular template, but their lattice parameters and molecular conformations change as compared to a silver surface. Individual molecules exhibit distinct appearances in scanning tunneling microscopy, consistent with site-dependent conformational variations identified by theory. Our calculations reveal that the competition between molecule–substrate interactions and conformational energy produces different compromise geometries, resulting in conformational frustration at the epitaxial interface. Crucially, constraining BTB into a planar geometry changes the preferred lateral adsorption positions, demonstrating that molecular flexibility directly modifies the energetic preferences for epitaxial registry. Comparison with more rigid molecular overlayers further supports this interpretation. Our findings reveal that molecular flexibility contributes to epitaxial matching by allowing unfavorable molecule–substrate interactions to be compensated by energetically favorable conformations while preserving the intermolecular network.

## Introduction

Hierarchical construction of complex materials from precisely arranged molecular building blocks is a central goal of nanotechnology.[1–6] In particular, extended two-dimensional (2D) nanostructures assembled from small organic molecules exhibit a wide range of favorable electronic, structural, magnetic, quantum and catalytic properties.[7–31] Advanced applications may require the integration of several functionalities, which can be achieved by combining different molecular building blocks into multilayer architectures.[6,32–36] Although the performance of such electronic devices often benefit from achieving well-ordered heteromolecular interfaces, the knowledge of heteromolecular systems[33–35,37–43] remains limited, and the key principles defining the epitaxial relationship at heteromolecular interfaces are incomplete.

Molecular epitaxy involves a complex interplay between intermolecular interactions, molecule-substrate interactions, and structural deformations. At interfaces comprising relatively rigid molecular networks, the epitaxy is mediated mainly through the mutual shift or twist of the layers[33,36,37,43] or distortion of otherwise preserved lattice[41]. Furthermore, Meissner et al. demonstrated that lateral displacements of otherwise rigid molecules can stabilize specific epitaxial orientations through competition between molecule-substrate and the energetic cost of intermolecular lattice deformation.[44] However, individual molecules may also exhibit intramolecular conformational degrees of freedom, allowing them to adapt to the underlying substrate without substantially modifying the intermolecular bonding network. Such adaptation can improve molecule-substrate interactions at the cost of conformational energy. While intramolecular flexibility has been shown to determine occupation[42] or chirality[38] in the top layer, the role of flexibility in establishing the epitaxial relationship remains insufficiently understood.

To address these challenges, we employ low-energy electron microscopy (LEEM) and diffraction (LEED),[45–47] scanning tunneling microscopy (STM), X-ray photoelectron spectroscopy (XPS),[48] and density functional theory (DFT) calculations to investigate the heteroepitaxy of conformationally flexible 1,3,5-tris(4-carboxyphenyl) benzene (BTB, Figure 1a). We choose a robust molecular δ-layer[49] as a substrate, which is composed of fully deprotonated BTB molecules on Ag(111). The δ-layer provides a stable molecular template that preserves the chemical identity of deposited BTB molecules, allowing us to investigate their conformational adaptation to the underlying substrate. We show that the binding motifs of self-assembled phases of BTB on Ag substrates are preserved on the molecular δ-layer, although their lattice parameters, orientations and molecular conformations are adapted to the new substrate. Our calculations identify two dominant contributions to the resulting epitaxial relationship. Intermolecular interactions largely define the molecular network, whereas the competition between molecule-substrate interactions and intramolecular conformational energy defines the energetics of epitaxial matching. This competition results in site-dependent conformational frustration, with individual BTB molecules adopting distinct compromise geometries within the epitaxial unit cell. This type of heteroepitaxy can thus generate periodic site differentiation within chemically uniform molecular layer, creating molecular moiré superlattices[36] without changing the molecular composition. Importantly, we demonstrate the role of conformational flexibility of the BTB molecule on preferred lateral positions by comparing the preferred adsorption positions of rigid molecules. Specifically, we show that BTB molecules constrained to planar geometries and more rigid HMTP molecules adopt different adsorption positions on the δ-layer due to the lack of the compensation mechanism. Our findings demonstrate that molecular flexibility contributes to epitaxial matching by enabling individual molecules to accommodate different local configurations while preserving the intermolecular network. Our work therefore contributes to understanding the driving forces of the epitaxial growth of heteromolecular systems, which is essential for incorporation of molecular functional layers into the heterointerfaces or with the traditional inorganic 2D materials.

## Results and Discussion

As a substrate for molecular heteroepitaxy, we have employed a compact layer of fully deprotonated BTB molecules on Ag(111), referred to as the δ-layer. Its structure was detailed in our previous paper.[49] In brief, the δ-layer homogeneously covers the entire substrate with minimum defects (Figure S1c in Supporting Information) and consists of two symmetry equivalent domains as documented in the color-coded diffraction pattern in Figure S1a and b in Supporting Information and dark-field image in Figure S1d in Supporting Information. Within the δ-layer, the BTB molecules appear as bright three-point star protrusions arranged in a hexagonal close-packed structure (Figure 1b) with $\begin{pmatrix} 5 & 1 \\ -1 & 4 \end{pmatrix}$ unit cell; we note that in this work, all superstructure unit cells are given in the matrix notation. The carboxylate (–COO) groups at the tips of the stars thereby point to the centers of neighboring molecules. The δ-layer is thermodynamically robust against mixing with organic semiconductors deposited on top and remains stable up to 235 °C, beyond which the decarboxylation occurs, resulting in a 2D polymeric network.[49] In the following experiments, we deposited the intact (protonated) BTB molecules onto the δ-layer (comprising fully deprotonated BTB). For clarity, unless explicitly stated otherwise, "BTB" refers only to the molecular layer formed on top of the δ-layer, and "δ-layer" refers to the substrate on which BTB growth occurs.

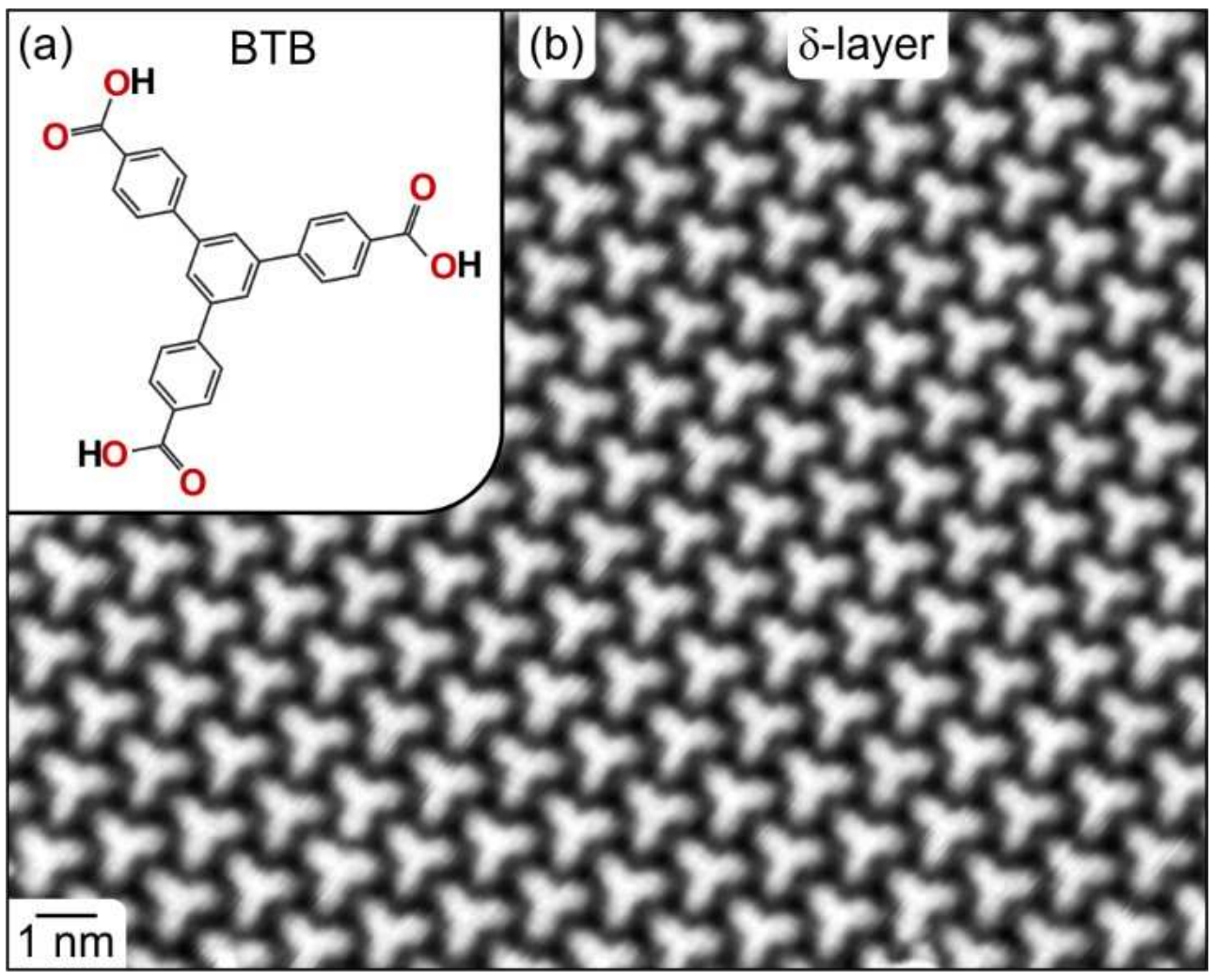

**Figure 1**: δ-layer on Ag(111). (a) Chemical structure of the intact BTB molecule. (b) STM image showing BTB molecules within the δ-layer; scanning parameters: 1.4 V, 50 pA.

**Structure of BTB on Ag(111) and on the δ-layer.** Starting with the δ-layer (Figure 2a), the deposition of BTB on top leads to the subsequent formation of two distinct phases, referred to as the honeycomb and stripe phases for their visual appearance; these phases can be distinguished by their diffraction patterns. The first one, honeycomb, appears after BTB deposition and persists up to 0.2 ML; i.e., 60 % of surface covered by BTB honeycomb phase, see Methods for definition of 1 ML. The diffraction pattern given in Figure 2b consists of two overlapping contributions: the intense one is associated with the δ-layer, and the weaker one with the BTB honeycomb phase. At this point, the majority of the surface is covered by the BTB honeycomb phase, which appears dark on the bright δ-layer in the associated bright-field LEEM image given in Figure 2b. With the additional deposition, diffraction from the stripe phase appears to coexist with that from the honeycomb phase. The diffraction pattern in Figure 2c becomes blurred except for the δ-layer diffraction. We stopped the deposition when the surface was almost completely covered by BTB molecules, as documented in the bright-field LEEM image in Figure 2c. Post-annealing at 70 °C for 30 s results in a sharp diffraction pattern (Figure 2d) that can be associated with the stripe phase. The subsequent annealing at 210–220 °C resulted in complete BTB desorption from the top layer, leaving the intact δ-layer as documented in Supporting Information, Figure S2. Both honeycomb and stripe phases were also observed directly on the Ag(111) substrate. In the following, we will provide the full structural characterization of these phases on both surfaces.

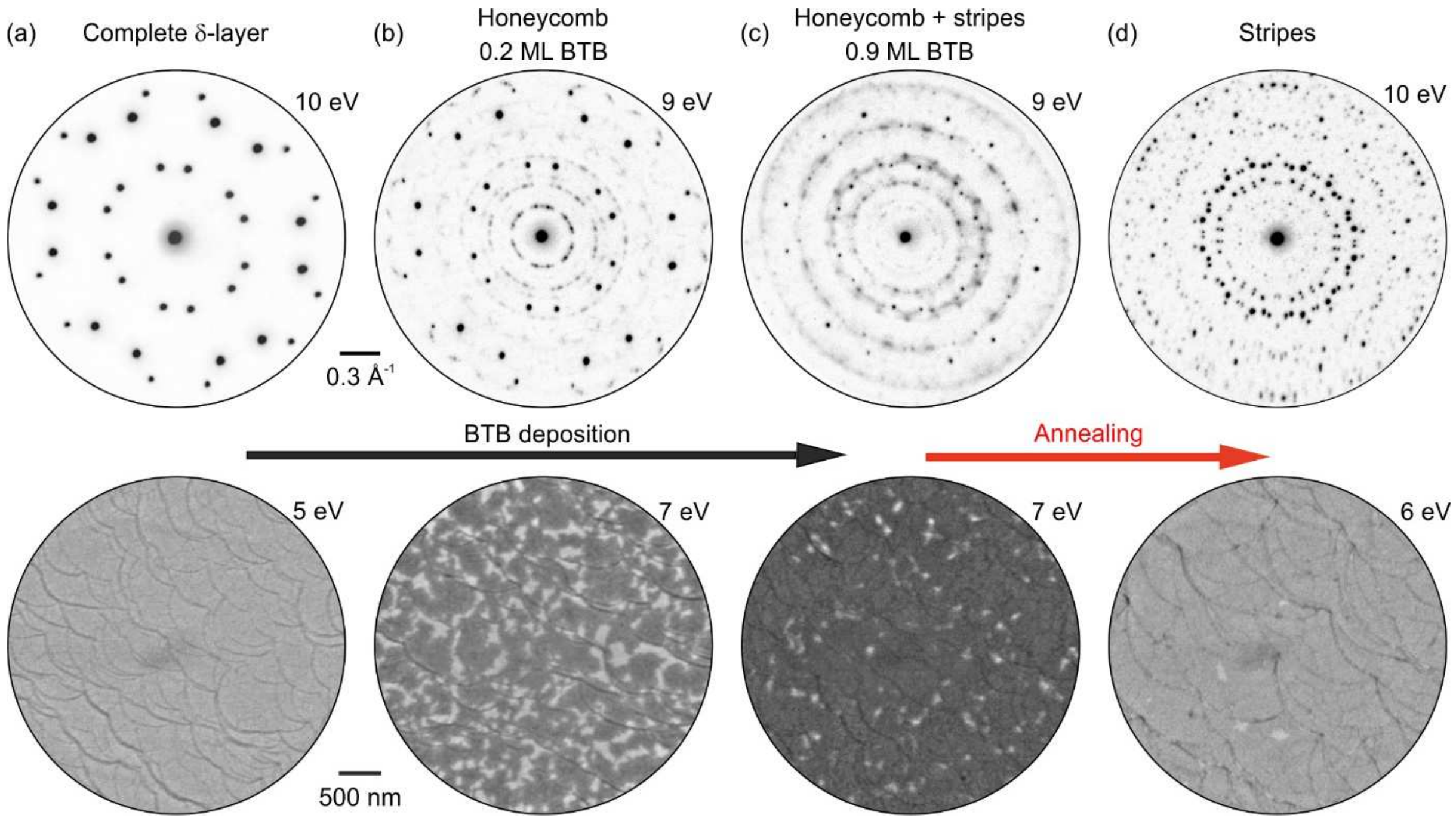

**Figure 2**: Evolution of diffraction patterns and LEEM bright-field images during BTB deposition on the δ-layer. (a) δ-layer on Ag(111). (b) Upon BTB deposition, first, the honeycomb phase is formed, which (c) partially transforms into the stripe phase for higher BTB coverages. The second BTB layer appears dark on the bright δ-layer. (d) After sample annealing at 70 °C, only the stripe phase is present on top of the δ-layer.

The structural analysis of BTB phases, both on the Ag(111) and the δ-layer, is given in Figure 3. In the honeycomb phase, the BTB molecules form an open structure with large hexagonal pores on both surfaces, as shown in the STM images and schematic representations in Figure 3a and 3g. On the Ag(111) substrate, the diffraction pattern associated with the honeycomb phase (Figure 3b) shows two overlapping hexagonal diffraction patterns associated with two symmetry-equivalent domains. The diffraction model in Figure 3c provides the unit cell with respect to the Ag(111) substrate of $\begin{pmatrix} 25 & 10 \\ -10 & 15 \end{pmatrix}$. On the δ-layer, the structure of the honeycomb phase is similar. The STM image in Figure 3g shows a well-resolved honeycomb phase together with the underlying δ-layer; this enables us to establish the matching of the honeycomb phase with the δ-layer as schematically shown in Figure 3g. Modeling of the diffraction pattern (Figures 3h and j) gives $\begin{pmatrix} 5 & -2 \\ 2 & 7 \end{pmatrix}$ unit cell with respect to the δ-layer and $\begin{pmatrix} 27 & -3 \\ 3 & 30 \end{pmatrix}$ to the Ag(111) substrate. The individual steps in modeling the honeycomb/δ-layer, including a detailed description of the diffraction pattern, are shown in Figure S3 in Supporting Information. The sizes of unit cells are summarized in Tables S1 and S2 in Supporting Information and discussed in detail below.

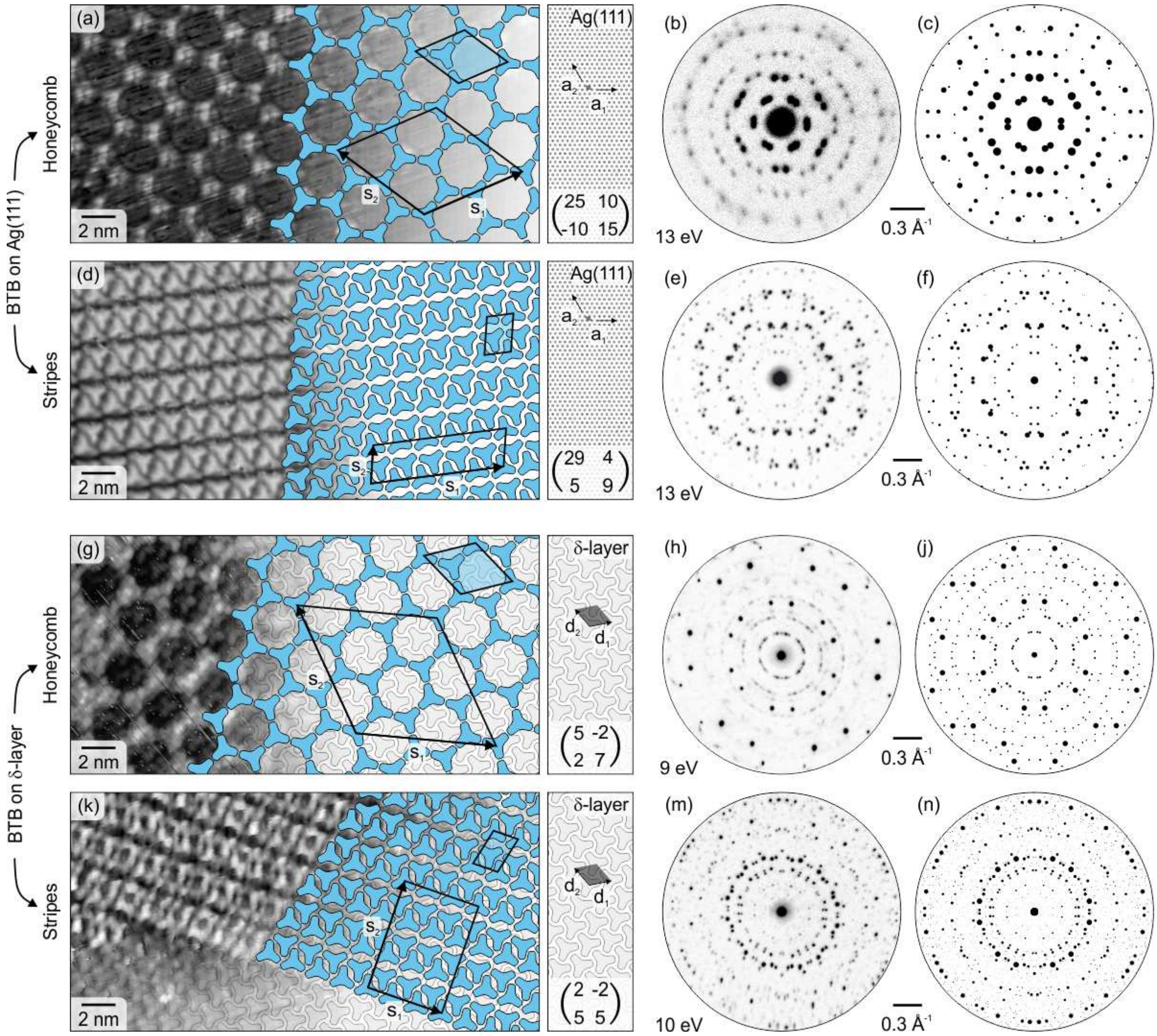

**Figure 3**: Structural analysis of BTB honeycomb and stripe phases. (a) STM – schematic representation, (b) diffraction, and (c) modeled diffraction of BTB honeycomb phase on Ag(111). (d) STM – schematic representation, (e) diffraction, and (f) modeled diffraction of BTB stripe phase on Ag(111). The model reproduces the principal stripe-phase reflections, while several weaker spots in (e) remain unassigned and may originate from minor coexisting structures or contaminants. (g) STM – schematic representation, (h) diffraction, and (j) modeled diffraction of BTB honeycomb phase on δ-layer. (k) STM – schematic representation, (m) diffraction, and (n) modeled diffraction of BTB stripe phase on δ-layer. The matrix notation of the commensurate unit cells (larger rhombus) with respect to the underlying surface (Ag or δ-layer) is given in the right-hand side panels of (b), (d), (g), and (k); the apparent unit cells are marked with a smaller parallelogram here. Scanning parameters: (a) 1.50 V, 50 pA, (d) 1.20 V, 50 pA, (g) 0.50 V, 50 pA, (k) 0.48 V, 50 pA.

Now we turn our attention to the second phase. In the stripe phase, the BTB molecules form a compact structure, as shown in the STM image in Figure 3d and in more detail in Figure 5a. The building up of the stripe phase from the BTB dimer is illustrated in Figure S5 in Supporting Information. In contrast to the honeycomb phase, in which the BTB molecules were bound together by complementary hydrogen bonds between two carboxyl groups, the stripe phase features an additional hydrogen bonding motif, i.e., a single hydrogen bond of carboxylic groups with the neighboring BTB molecule. Similar motifs were observed on the Au(111) surface.[50] However, on Au, several motifs with distinct arrangements of the basic dimer were present both on liquid-solid[51] and vacuum solid interface (our results, see Figure S6 in Supporting Information). Calculated gas-phase structures and a discussion of their stability are provided in Supporting Information, Section 7. In brief, while the stripe phase features only one strong complementary carboxyl–carboxyl bond, the energy is compensated by 2 additional (weaker) bonds with the neighboring molecules, which makes both phases close in energy. However, at high coverages, the stripe phase is observed to be dominant, as it features a higher molecular density per unit area (see Table S2 in Supporting Information).

The diffraction pattern of the stripe phase on Ag(111) shown in Figure 3e is quite complex, as this phase displays six symmetry-equivalent domains. The diffraction model in Figure 3f provides the $\begin{pmatrix} 29 & 4 \\ 5 & 9 \end{pmatrix}$ unit cell with respect to the Ag(111). On the δ-layer, although imaging of individual BTB molecules is limited due to their structural adaptation to the substrate (Figure 3k), the periodic motifs remain clearly recognizable. The overview image in Supporting Information, Figure S8, shows the extended unit cell commensurate with the underlying δ-layer.

The structure of the stripe phase is similar to that on Ag(111); however, the individual stripes on the δ-layer appear less separated, resulting in higher molecular density (see Table S2 in Supporting Information). Modeling the diffraction pattern (Figures 3m and n) gives $\begin{pmatrix} 2 & -2 \\ 5 & 5 \end{pmatrix}$ unit cell with respect to the δ-layer and $\begin{pmatrix} 12 & -6 \\ 20 & 25 \end{pmatrix}$ to Ag(111).

**XPS analysis of BTB on Ag(111) and the δ-layer.** Finaly, we will focus on the chemical state of BTB. The photoelectron O 1s spectrum of δ-layer (Figure 4, top) shows a single peak associated with fully deprotonated BTB molecules. After the BTB deposition of 0.2 ML on top of δ-layer, two peaks associated with deposited BTB appear, whereas the δ-layer peak does not show any significant changes. This is observed for BTB coverages up to 1 ML. For comparison, the photoelectron spectrum of the BTB deposited directly on the Ag(111) is shown in Figure 4 (marked as α/Ag(111)); it features two peaks associated with hydroxyl (C–<u>O</u>H) and carbonyl (C=<u>O</u>) oxygens of the carboxylic group (–COOH), indicating the intact BTB. In comparison with BTB/Ag, the peaks associated with BTB on the δ-layer are only rigidly shifted to lower binding energies due to the change of substrate work function (WF)[52] between Ag(111) and the δ-layer, but keep their mutual binding energy shift and ratio of intensities. Hence, the photoelectron spectra indicate that the BTB deposited on top of the δ-layer stays intact.

Both honeycomb and stripe phases formed on top of the δ-layer comprise only intact (protonated) molecules, representing thus two polymorphs of the intact phase. These phases were also observed directly on the Ag(111) substrate, although the stripe phase was previously tentatively assigned to feature 1/3 of the deprotonated carboxyl groups.[53] In contrast, our XPS data for as-deposited BTB indicate a very low to zero degree of deprotonation of carboxyl groups.[49] Nevertheless, initial phases comprising carboxylic acids may accommodate a certain number of deprotonated molecules[54,55] without significant structural change. Additionally, for degrees of deprotonation up to 2/3, we observe the compressed stripe phase (see Figure S9 in Supporting Information), where the stripes are closer together, indicating that relatively small structural changes are necessary to accommodate even a significant degree of deprotonation within the stripe phases. Based on these results, we propose that the stripe phase on Ag(111) is fully protonated and can accommodate a limited number of partially deprotonated BTB molecules.

Annealing of the sample leads to BTB desorption from the second layer, leaving the δ-layer intact. The desorption of BTB from the top of the δ-layer takes place at temperatures as high as 210–220 °C, which shows a considerably strong interaction with the δ-layer. We note that the BTB desorption temperature from the δ-layer is only ~ 5 °C below the decarboxylation threshold above which the carboxylate groups are cleaved from the phenyl rings.

The work function shows a negligible decrease of 0.015 eV upon deposition of 0.2 of ML of BTB on the δ-layer and remains constant with further deposition up to 0.6 ML. We previously established a linear trend between the measured WF and the core-level positions of the C 1s associated with phenyl rings of aromatic carboxylic acids.[49,52] The updated graphs are given in Figure S10 in Supporting Information. The binding energy of 284.3 eV measured for the phenyl-related component of the C 1s in the δ-layer fits this trend. Concerning the BTB on top of the δ-layer, the position of the phenyl component of the C 1s peak is 284.8 eV, again fitting the linear trend established between the substrate WF and the position of the C 1s peak in the BTB.[10] The rigid shift of both O 1s and C 1s peaks of BTB deposited on top of δ-layer with the change of substrate work function is a signature of vacuum-level alignment and weak electronic coupling typical for adsorption dominated by van der Waals interaction.[49,52] If we consider the BTB on the δ-layer, a minimal shift in the work function of the system takes place. This presents a significant advantage over metallic substrates, where the adsorption of molecules can significantly change the work function due to pushback effects. Together with its temperature stability and robustness against mixing with other molecules, this makes the δ-layer a robust reference substrate for determining properties of molecular interfaces.

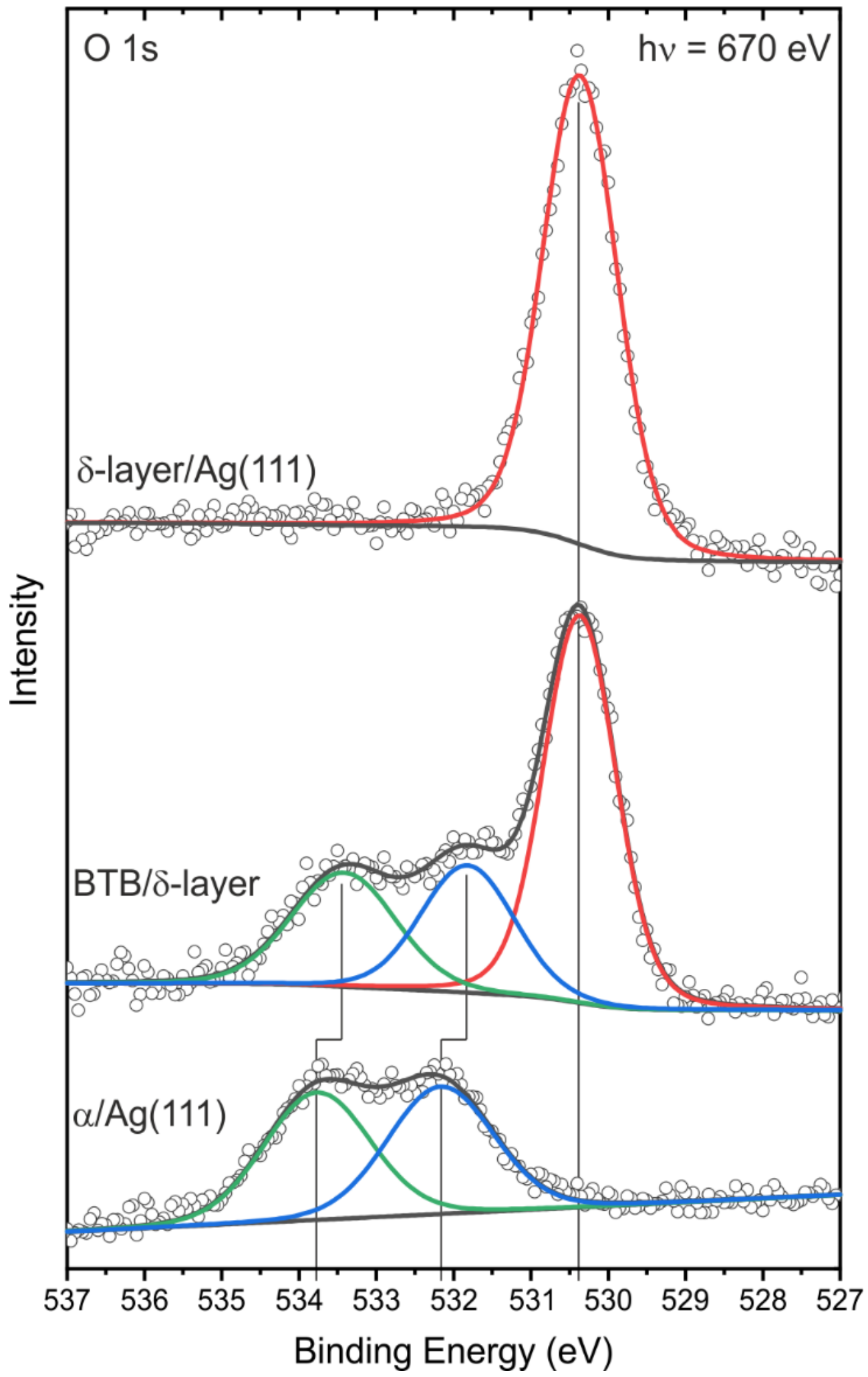

**Figure 4**: Photoelectron spectra of the O 1s region measured with synchrotron radiation of 670 eV energy on the δ-layer on Ag(111), after deposition of 0.6 ML of BTB on the δ-layer. The spectrum of BTB deposited on the Ag(111) surface held at 20 °C is shown for comparison. The peak-fitting parameters are given in Table S4 in Supporting Information.

**Origin of the distinct BTB unit cells on Ag(111) and the δ-layer.** BTB phases on top of the δ-layer resemble those deposited directly on the Ag(111) substrate. Both have well-defined orientations with respect to the δ-layer substrate. The BTB superlattices do not have a simple relation with the substrate, but they are commensurate with a larger periodicity. Substrate is an integral part of self-assembly; therefore, the key aspects of multilayer self-assembly are the stability and robustness of the substrate, which were discussed in our previous work.[49]

Both honeycomb and stripe phases are compressed when translated from Ag(111) to the δ-layer. The lattice parameter of the honeycomb phase decreased from 31.49 Å on Ag(111) to 31.25 Å (δ-layer), i.e., by –0.76%, while the angle between unit cell vectors remains 120°, as discussed in more detail in Tables S1 and S2 in Supporting Information. Lattice parameters of the stripe phase changed from 22.57 × 15.73 Å$^2$ on Ag (111) to 22.65 × 15.29 Å$^2$ on δ-layer, and the angle between unit cell vectors changed from 79° to 77°. Thus, one lattice parameter is expanded by +0.35% and the second one is compressed by –2.8%. Moreover, the translation of both phases onto the δ-layer is accompanied by changes in the appearance of BTB molecules in the STM images and variability in the apparent heights of BTB parts (see Figure 3), pointing to a tilt or bending of the phenyl rings.

We have performed DFT calculations to explain structural changes observed in the experiment. Already calculations on unsupported phases, i.e., in the absence of the substrate, explain the difference in the unit cell of both phases. We have assumed two configurations for both phases: in the first, all BTB molecules are in a planar configuration, and in the second, the molecules are non-planar, i.e., in conformation with phenyl rings rotated as shown in Figure 5b. In both phases, the non-planar configuration is ~250 meV per BTB molecule more stable than the planar one, which we attribute to the optimization of the distances between (partially positively charged) hydrogen atoms in neighboring phenyl rings of the BTB molecule. Importantly, changes in the calculated lattice parameters are in perfect agreement with experiment: the lattice parameter of the honeycomb phase is changed upon tilting by –0.94%, and the lattice parameters of the stripe phase are changed by +0.48% and –2.85% for both lattice vectors, respectively. Our DFT model qualitatively explains a 3 times larger change in the second lattice vector of the stripe phase relative to the honeycomb lattice vector in the following way: bonding inside the stripe involves hydrogen atoms of phenyl rings, which, upon rotation, come closer to the neighboring BTB molecule; this results in the change of the unit cell shown in Figure 5a. On the other hand, bonding through the carboxylic groups, present between the stripes and in the honeycomb phase, is less sensitive to the tilt of the phenyl rings.

Therefore, our results point to the following scenario: molecular phases adsorbed on Ag(111) are planarized by van der Waals interactions with the substrate, leading to lateral extension. When molecular phases are translated onto the δ-layer, van der Waals interactions weaken, leading to a structure similar to the non-planar configuration, which, in turn, changes the lattice parameters. The non-planar configuration also explains the different STM appearance between the molecular phases adsorbed on Ag(111) and on the δ-layer in Figure 4. δ-layer thus largely preserves the optimal gas phase configuration featuring the tilted molecules and the unit cell size.

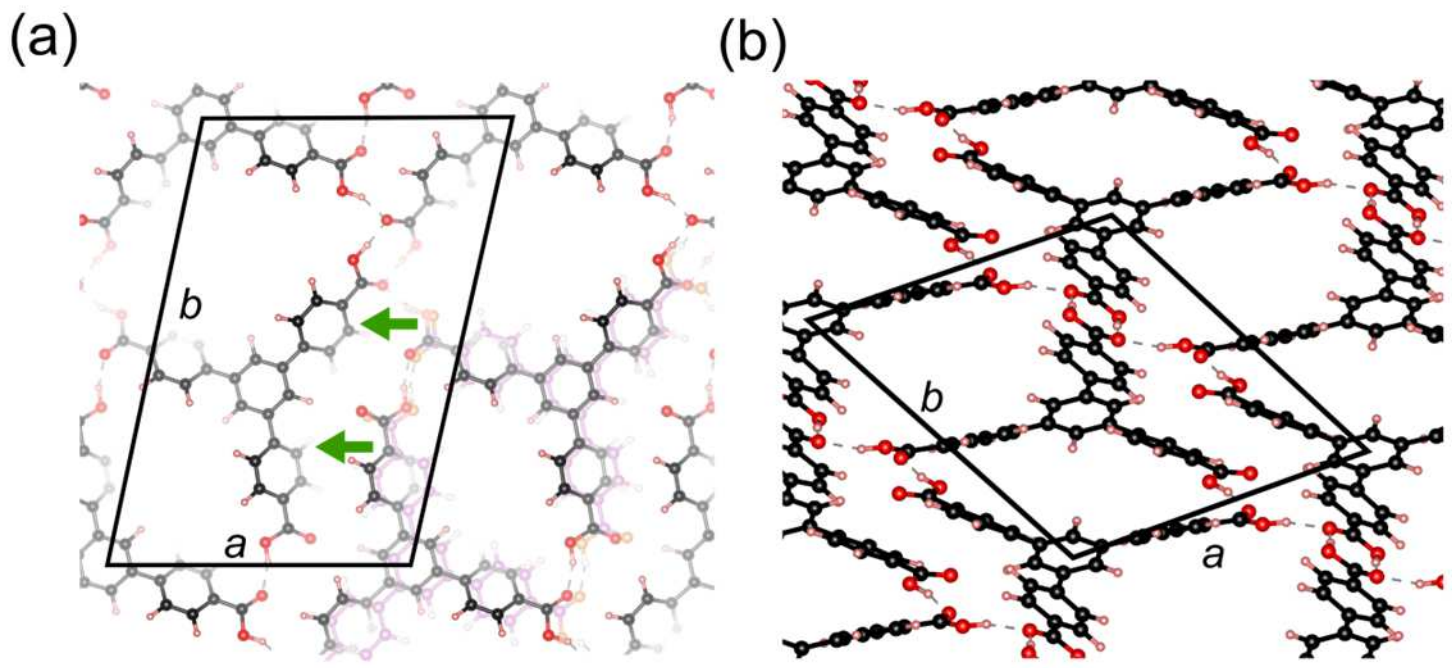

**Figure 5:** Models of unsupported stripe phase**.** (a) top and (b) perspective view on the unsupported, corrugated stripe phase calculated with DFT. In panel (a), the corrugated phase colored in black is overlaid with the planar configuration drawn in violet. Compression along the lattice vector $a$ induced by corrugation is marked with green arrows. The models of unsupported honeycomb phase are shown in Figure S11 in Supporting Information.

**Conformationally frustrated epitaxy of BTB on the δ-layer.** The preceding calculations explain how conformational relaxation influences the lattice parameters of BTB phases on the δ-layer. However, they do not address another important aspect of epitaxy: the preferred lateral alignment of the molecular overlayer with the underlying substrate. Within the extended epitaxial unit cell, individual BTB molecules occupy different local adsorption environments. Therefore, we performed DFT calculations to uncover the driving forces governing the heteroepitaxy of the BTB overlayer on the δ-layer. Figure 6a-b shows the decomposition of the relative stability of the calculated structures into three physically distinct contributions: (i) the interaction of BTB with the underlying δ-layer, denoted as $E_{\mathrm{int}}$; (ii) the conformation energy of the BTB molecule ($E_{\mathrm{conf}}$); and (iii), the intermolecular interaction within the BTB overlayer ($E_{\mathrm{BTB-BTB}}$). The interaction energy between an individual BTB molecule and the δ-layer ($E_{\mathrm{int}}$) is defined as

$$E_{\mathrm{int}} = E_{\mathrm{BTB/\delta-layer}} - E_{\mathrm{BTB}} - E_{\mathrm{\delta-layer}},$$

where $E_{\mathrm{BTB}/\delta-\mathrm{layer}}$ is the energy of an isolated BTB molecule adsorbed on the δ-layer in a position and conformation obtained from the complete overlayer, $E_{\mathrm{BTB}}$ is the energy of the same BTB molecule in the gas phase with the same conformation, and $E_{\delta-\mathrm{layer}}$ is the energy of the isolated δ-layer. The molecular conformation energy ($E_{\mathrm{conf}}$) accounts for the energetic cost of deforming the BTB molecule from its optimized gas-phase geometry into the configuration adopted in the overlayer. Finally, intermolecular interaction between the neighboring BTB molecules in the overlayer, $E_{\mathrm{BTB-BTB}}$, is dominated by hydrogen bonding between neighboring carboxylic groups, as already discussed for the gas-phase models. We define this energy as

$$E_{\mathrm{BTB-BTB}} = (E_{\mathrm{overlayer}} - \sum E_{\mathrm{conf}})/N,$$

where $E_{\mathrm{overlayer}}$ is the total energy of the BTB overlayer composed of $N$ molecules. Importantly, when summing all three contributions for all BTB molecules in the calculated overlayer, we obtain the total energy of a system, a decisive parameter for stability.

In the following, we show a clear separation between the interactions responsible for defining the molecular network and those that select its epitaxial registry. Overall, we optimized 30 configurations of the BTB stripe phase adsorbed on the δ-layer, each containing 8 BTB molecules in the overlayer. The initial structures were generated by distinct lateral translations of the complete stripe overlayer relative to the δ-layer. Importantly, the BTB molecules were initially fully planarized to allow them to adapt freely to the underlying δ-layer during geometry optimization. Across all 30 configurations, Figure 6c–i shows the correlations between total energy and specific energy contributions, as well as their combinations, averaged over all 8 BTB molecules in the overlayer unit cell. Note that such analysis cannot be performed at the single-molecule level, as the collective balance among the BTB molecules within the calculated supercell determines the observed structure: the BTB overlayer shows the best compromise of adsorption energies of BTB molecules, with their positions dictated by the molecular structure of the overlayer.

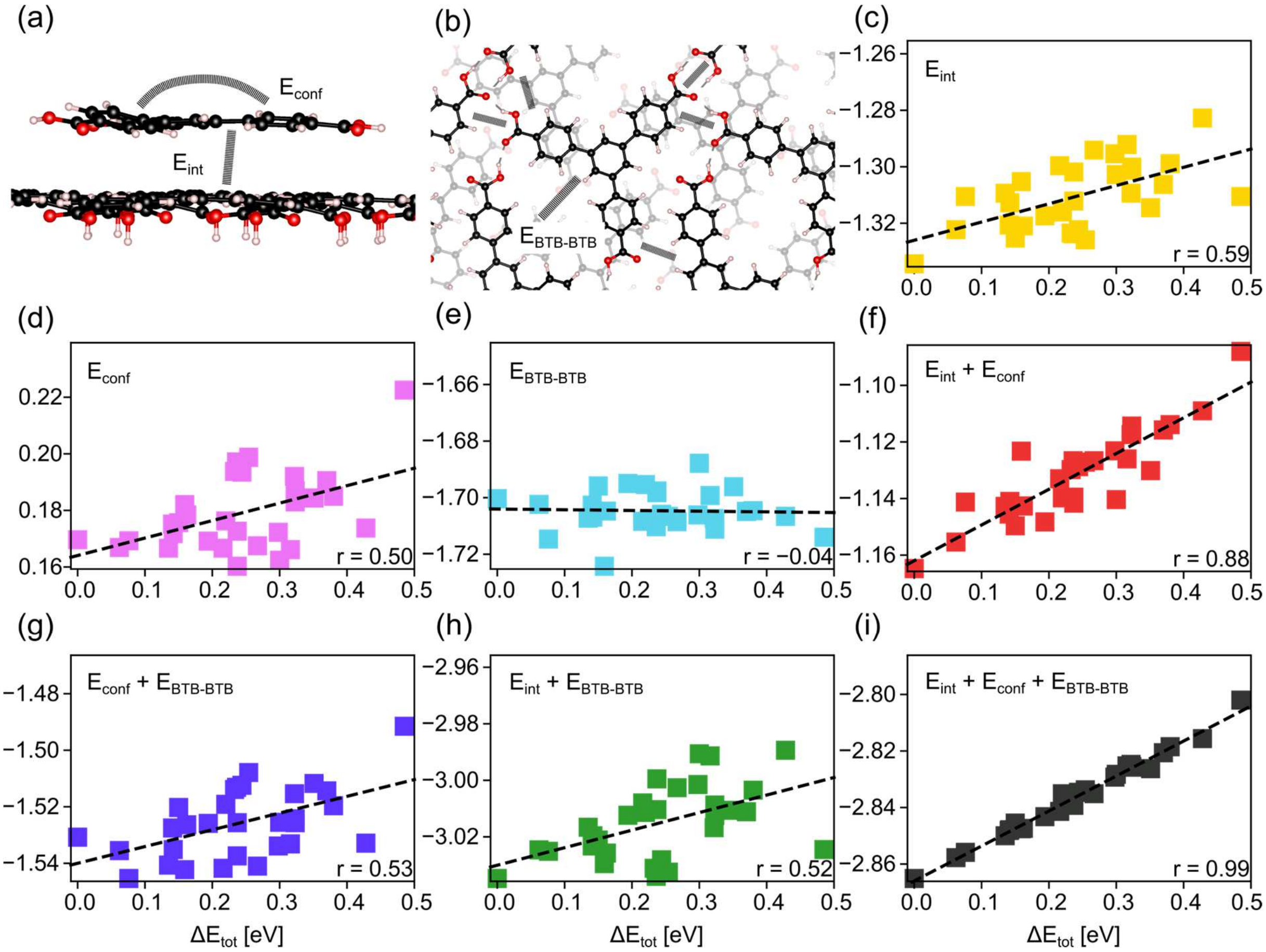

**Figure 6:** Correlation between the three interactions in the BTB overlayer on the δ-layer and the total energy of the system plotted on the *x*-axis in all panels. (a) Side and (b) top view of the BTB on the δ-layer, with the interactions marked with black dashed lines: $E_{\mathrm{int}}$ is the interaction energy between adsorbed molecules and the substrate, $E_{\mathrm{BTB-BTB}}$ energy between the BTB molecules, and $E_{\mathrm{conf}}$ energy of their conformational change with respect to the optimal conformation. The *y*-axis in panels (c-i) shows changes in these energies (c, d, e) and their combinations (f, g, h, i) in eV as a function of the change of total energy of the system. Calculated correlation coefficients (r) are shown at the bottom of each panel. All energy contributions are reported per BTB molecule; the presence of eight BTB molecules in the supercell thus partially averages out favorable and unfavorable local adsorption sites, which reduces the spread of the energy contributions.

Neither $E_{\mathrm{int}}$ nor $E_{\mathrm{conf}}$ alone shows a strong correlation with the total energy of the system, with correlation coefficients of $r$ = 0.50 and $r$ = 0.59, respectively, as shown in Figures 6c–d. Remarkably, $E_{\mathrm{BTB-BTB}}$ shows essentially no correlation with the stability of the interaction energy $E_{\mathrm{int}}$. Thus, the intermolecular interactions are crucial for preserving the hydrogen-bonded BTB network but do not affect the epitaxial relationship and optimal position of the BTB phases with respect to the δ-layer. In contrast, combining $E_{\mathrm{int}}$ and $E_{\mathrm{conf}}$ results in a stronger correlation with the total energy ($r$ = 0.88), as shown in Figure 6f. Their sum can thus be interpreted as the effective adsorption energy of an individual BTB molecule in the overlayer, including both BTB interaction with the δ-layer and the energetic cost required to adopt its configuration. The strong correlation, therefore, identifies the balance between interfacial interactions and molecular deformation as the primary factor governing the epitaxy of the BTB overlayer. When all three energy contributions are combined, a nearly perfect correlation with the total energy is obtained Figure 6i, confirming that numerical errors introduced by the decomposition are negligible.

To demonstrate why the first two interactions are important, we will discuss representative examples of isolated BTB molecules adsorbed on the δ-layer. We start from the calculated models in the 50 meV interval of the least optimal $E_{\mathrm{int}}$. In this particular case, the BTB molecules show direct overlap of phenyl rings with those of the molecules below, which form the δ-layer. The instability of such a configuration is expected, as the face-to-face position of phenyl rings should be avoided due to Pauli repulsion.[56] Conversely, models with optimal $E_{\mathrm{int}}$ preferentially adsorb either above the empty regions of the δ-layer or in configurations where part of their phenyl rings approaches the laterally offset geometry favorable for π-π stacking.

Molecular flexibility, however, prevents the interaction energy from independently determining the overall stability. A BTB molecule can improve its interaction with the δ-layer by changing its optimal conformation, but such a change results in an energetic penalty relative to the gas-phase configuration. Conversely, retaining a more favorable configuration can force parts of the BTB molecule into suboptimal positions with respect to the underlying δ-layer. Therefore, $E_{\mathrm{int}}$ and $E_{\mathrm{conf}}$ can compensate each other thanks to the flexibility of the BTB molecule. This compensation between the two contributions is most clearly demonstrated in Figure 7, where $E_{\mathrm{int}}$ is plotted against $E_{\mathrm{conf}}$ for all 240 individual BTB molecules calculated in our dataset. The diagonal lines correspond to constant adsorption energies ($E_{\mathrm{ads}} = E_{\mathrm{int}} + E_{\mathrm{conf}}$). The distribution of configurations along these lines (Figure 7a) shows that similar adsorption energies can result from substantially different contributions of $E_{\mathrm{int}}$ and $E_{\mathrm{conf}}$. The two configurations highlighted in Figure 7a and shown in Figures 7b and c provide a particularly clear example of this compensation. One has the most favorable $E_{\mathrm{int}}$ in the entire dataset, while the second one has the most favorable $E_{\mathrm{conf}}$. Their interaction energies differ by 101 meV, and their conformational energies differ by 107 meV. Yet, their resulting difference in the adsorption energy is only 6 meV. The less favorable interaction energy in one of these configurations can be related to a partial overlap of phenyl rings with the underlying layer, as shown in Figure 7b, whereas this overlap is avoided in the configuration with more favorable $E_{\mathrm{int}}$ (Figure 7c). Similar adsorption energies can therefore arise from different trade-offs between interfacial interactions and molecular conformation. Thus, the balance between $E_{\mathrm{int}}$ and $E_{\mathrm{conf}}$ is not selected universally for all molecules in the unit cell, but each molecule follows its own optimum, forming a conformationally frustrated system. In contrast, our further calculations show that the epitaxy and lateral positions of overlayers composed of rigid aromatic molecules with strong intermolecular bonding are governed primarily by the optimization of the interaction energy $E_{\mathrm{int}}$, as discussed later.

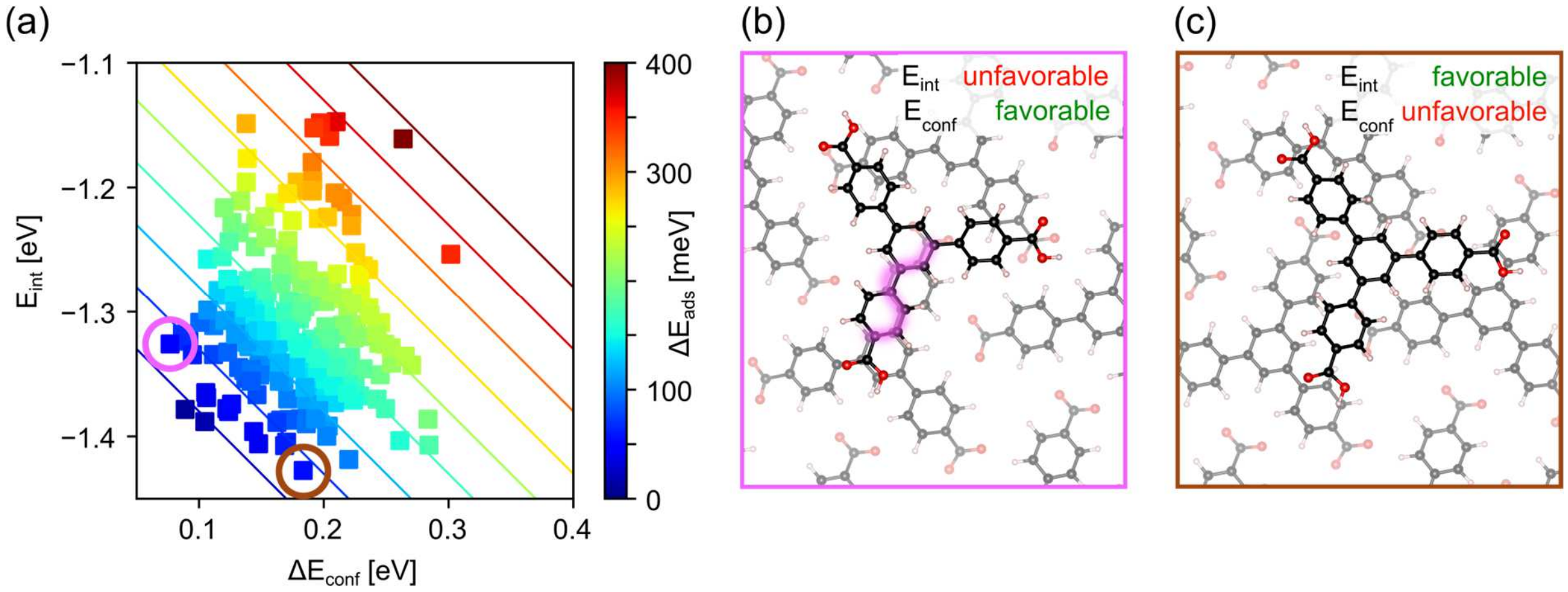

**Figure 7**: Compensation between interaction energy $E_{\mathrm{int}}$ and conformation energy $E_{\mathrm{conf}}$. Panel (a) shows the relative values of $E_{\mathrm{conf}}$ plotted against $E_{\mathrm{int}}$ for all 240 individual BTB configurations, with the points colored according to relative adsorption energy $E_{\mathrm{ads}}$. The diagonal lines correspond to constant $E_{\mathrm{ads}}$ and show that a less favorable $E_{\mathrm{int}}$ can be compensated by more favorable $E_{\mathrm{conf}}$, and vice versa. Panels (b) and (c) show the two highlighted configurations, which differ by 101 meV in $E_{\mathrm{int}}$ and by 107 meV in $E_{\mathrm{conf}}$. Yet, their $E_{\mathrm{ads}}$ differs only by 6 meV. The configuration in panel (b) shows a partial overlap of phenyl rings with the underlying layer, highlighted in purple.

The resulting energy landscape contains a broad manifold of nearly degenerated structures. We identify 43 configurations within 100 meV of the most favorable $E_{\mathrm{ads}}$. These configurations represent different compromises between the interaction and conformational energies. Two limiting structural motifs occur in this ensemble: (i) molecules placed predominantly above the empty regions of the δ-layer generally achieve more favorable $E_{\mathrm{int}}$, but adopt a less favorable molecular conformation and therefore have higher $E_{\mathrm{conf}}$. A particular example of such a configuration is illustrated in Figure 7c. (ii) In contrast, a configuration with partial overlap between molecules in the two layers typically preserves a more favorable $E_{\mathrm{conf}}$ while being less optimal in $E_{\mathrm{int}}$. A typical example of such a configuration is shown in Figure 7b. This second scenario can be further divided into two subgroups. In the first, only a smaller part of the BTB molecule directly overlaps with the underlying molecule, while its larger part remains above the empty region of the δ-layer. In the second, a larger part of the molecule overlaps with the underlying BTB, but overlapping phenyl rings adopt a lateral position favorable for π-π stacking.

This interpretation is further supported by configurations with the least favorable $E_{\mathrm{ads}}$ (geometries colored in red in Figure 7a). These configurations do not exploit either the empty regions of the δ-layer, or a favorable π-π stacking position. Instead, all structures show a larger direct overlap with the underlying BTB molecules, resulting in unfavorable $E_{\mathrm{int}}$. In these particular cases, compensation with $E_{\mathrm{conf}}$ is limited due to the steric hindrance with the underlying molecules. As a result, geometries within 100 meV of the least favorable $E_{\mathrm{ads}}$ are strongly planarized, and therefore also unfavorable in $E_{\mathrm{conf}}$, as shown in Figure S12 in Supporting Information. The compensation mechanism is again stronger in geometries within the 100 meV to 200 meV interval of the least stable structures (orange to light green points in Figure 7a). Two distinct classes of geometries can be identified within this interval: (i) geometries with a significant overlap with the underlying BTB molecules, which are favorable in $E_{\mathrm{conf}}$, but unfavorable in $E_{\mathrm{int}}$. (ii) strongly planarized geometries (unfavorable $E_{\mathrm{conf}}$) situated in positions close to empty regions of the δ-layer, or in a favorable π-π stacking position (favorable $E_{\mathrm{int}}$).

To test whether this conformational adaptation stabilizes the heteroepitaxial structure, we performed an alternative set of optimized geometries starting from the optimized gas-phase network placed on the δ-layer. The intermolecular interaction ($E_{\mathrm{BTB-BTB}}$) defines also in this case the mutual position of BTB molecules, allowing only for their conformational changes, because the magnitude of $E_{\mathrm{BTB-BTB}}$ is larger than that of $E_{\mathrm{int}}$ (Figures 6c, e). However, when starting from the gas phase configuration, the BTB molecules are trapped in local minima on the potential energy surface, and they cannot fully adopt the optimal geometry relative to the underlying layer. Therefore, the compensation mechanism is strongly suppressed, as further discussed in Supplementary Information, Section 14. Suppressing the full adaptation to the δ-layer produces a clear energetic penalty. As expected, the average $E_{\mathrm{conf}}$ is relative to the original dataset 20 meV more favorable and $E_{\mathrm{BTB-BTB}}$ is improved by approximately 10 meV per BTB. However, these gains are outweighed by 60 meV less favorable interaction energy. In summary, suppression of the compensation mechanism leads to the average energy penalty of ~30 meV per BTB molecule. This clearly quantifies the trade-off among these three contributions, which together determine the overall stability of the system.

The change in molecular uniformity between Ag(111) and δ-layer is experimentally observed for BTB phases. The STM images of BTB stripe phases on Ag(111) and the δ-layer in Figure 8a-b show a remarkable difference in the appearance of individual BTB molecules. On Ag(111), all the BTB molecules appear very similarly, whereas on the δ-layer, each BTB within the extended unit cell shows a very distinct appearance due to its different adaptation to the underlying δ-layer. A similar situation is for the honeycomb phase in Figure 8c. As the BTB molecules have a considerable band gap, this is a structural effect reflecting variations in BTB molecular orientation, tilt, or bending that accompany small changes in electronic structure, consistent with the results of DFT computations. On the other hand, while the molecular conformation varies within the unit cell, it periodically repeats in the commensurate lattice, as shown in Figure S8 in Supporting Information. The heteroepitaxy thus creates a moiré superlattice in which the molecular properties periodically vary. Variation in molecular appearance renders the BTB molecular phases non-uniform at the molecular level, thereby reducing their symmetry, i.e., lifting both mirror and rotational symmetries.

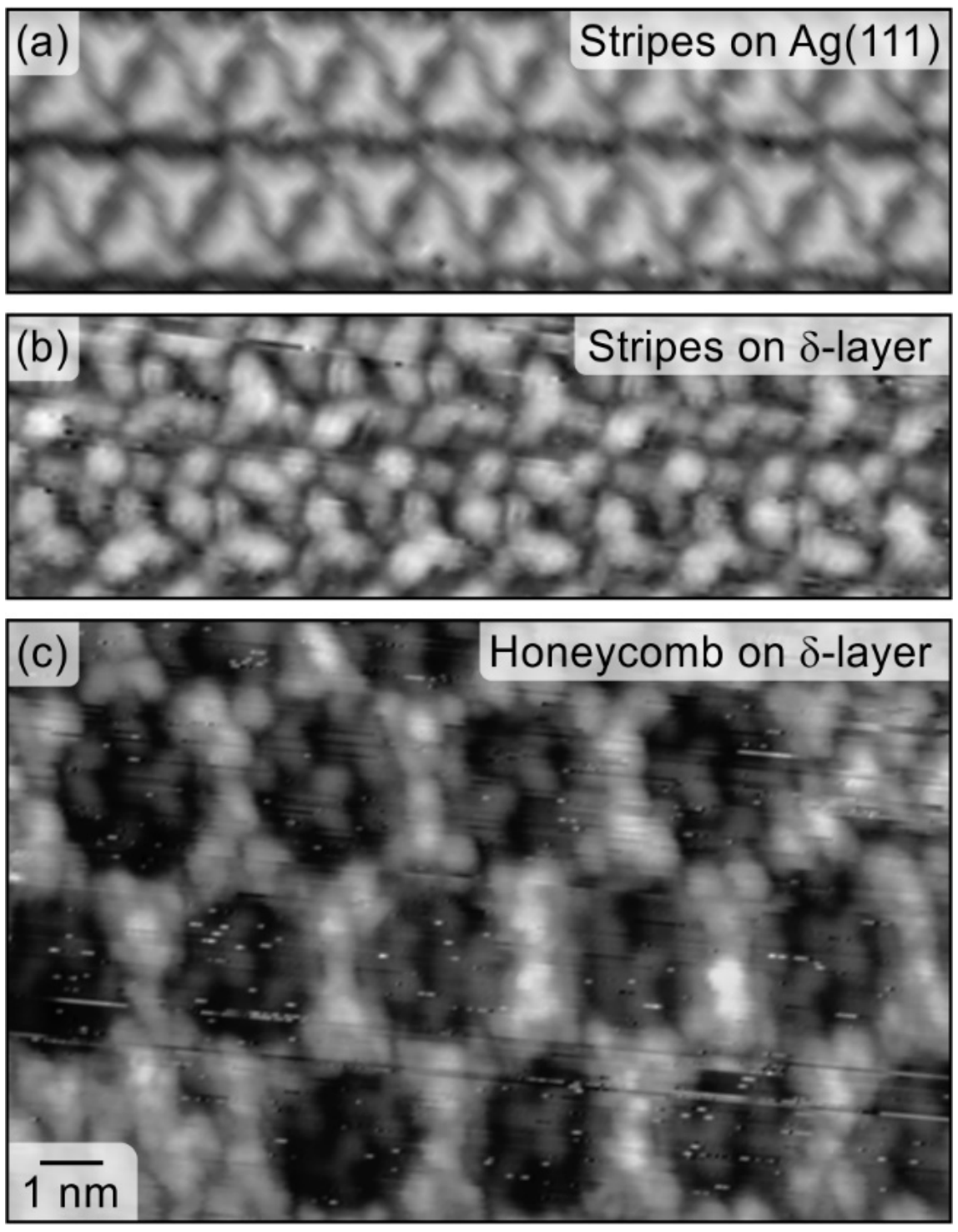

**Figure 8**: High-resolution STM images of BTB phases. (a) STM image of the stripe phase on Ag(111). (b) STM image of the stripe phase on the δ-layer. Note the periodic appearance of every third BTB in the horizontal direction in each line. (c) STM image of the honeycomb phase on the δ-layer. Scanning parameters: (a) 1.2 V, 50 pA, (b) 0.5 V, 50 pA, (c) 0.5 V, 50 pA.

**Generalization of the model: HMTP on the Ag(111) and δ-layer.** The theoretical framework given above considered the case in which the intermolecular interaction $E_{\mathrm{BTB-BTB}}$ defined the framework shape and completion of the layer-substrate $E_{\mathrm{int}}$ and conformation energy $E_{\mathrm{conf}}$ provided the epitaxy condition. To assess the limiting case of the presented energy partitioning, we have employed 2,3,6,7,10,11-hexamethoxytriphenylene (HMTP, Figure S15a in Supporting Information)[57,58] molecules with reduced conformational flexibility and weaker intermolecular interactions. The HMTP lattice matches that of the δ-layer within 1%, allowing us to employ a minimal supercell containing one HMTP molecule and one BTB molecule. Our theoretical framework predicts that reducing the molecular flexibility suppresses the competition between $E_{\mathrm{int}}$ and $E_{\mathrm{conf}}$. Consequently, the lateral position of a rigid molecular network should be defined solely by its interaction with the underlying layer, whereas intermolecular interactions determine the mutual arrangement of the molecules within the network, as illustrated in Figure 9 and detailed in Supporting Information, Section 15. The total energy correlates almost perfectly with $E_{\mathrm{int}}$ ($r$ = 0.98), while $E_{\mathrm{conf}}$ spans only a narrow energy range and shows much weaker correlation ($r$ = -0.36). Similarly, changes in the intermolecular energy $E_{\mathrm{HMTP-HMTP}}$ is nearly constant among the different configurations, consistent with its behavior in the BTB overlayer. The absence of a broad distribution of $E_{\mathrm{int}}$ across $E_{\mathrm{conf}}$ further confirms that the compensation mechanism identified for the flexible BTB overlayer is essentially absent for HMTP. The lateral position of the HMTP network is therefore selected by maximizing the attractive interaction with the δ-layer. Consistent with the BTB overlayer, the calculated potential energy surface (Figure S14) shows that these preferred positions correspond to the empty regions of the underlying layer. This theoretical prediction matches experiments detailed in Supporting Information, Section 16: the HMTP grows epitaxially on the δ-layer with the simple superstructure unit cell matching to the underlying δ-layer.

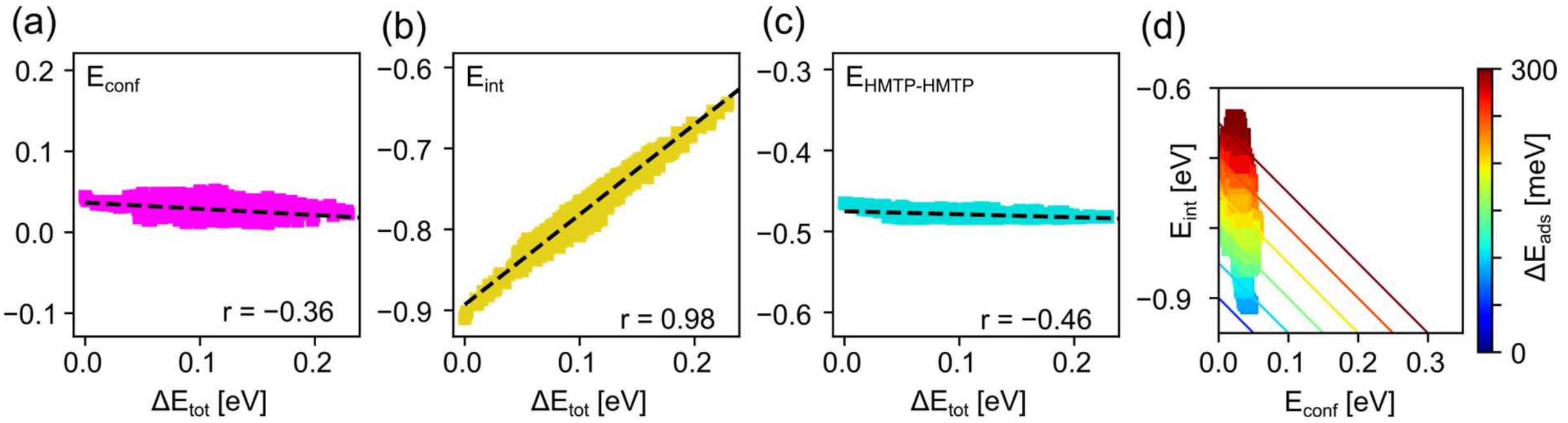

**Figure 9:** Correlation analysis for the HMTP molecule on the δ-layer. Correlation between the total energy of the system $\Delta E_{\mathrm{tot}}$, and: (a) conformational energy $E_{\mathrm{conf}}$, (b) interaction energy with the underlying δ-layer $E_{\mathrm{int}}$, and (c) intermolecular interaction energy $E_{\mathrm{HMTP-HMTP}}$. Dashed lines represent the linear fit of acquired data, and $r$ denotes the correlation coefficient. Panel (d) shows the relationship between the interaction energy $E_{\mathrm{int}}$ and the conformation energy $E_{\mathrm{conf}}$. The narrow range of $E_{\mathrm{conf}}$ and the strong correlation between $E_{\mathrm{int}}$ and $\Delta E_{\mathrm{tot}}$ show that the compensation mechanism presented for the HMTP overlayer is strongly suppressed. The relative stability between different lateral positions is determined by the interaction energy $E_{\mathrm{int}}$. Note that the larger spread on the $y$-axis in panels (a–c), compared with the BTB overlayer analysis in Figure 6, results from the absence of averaging over multiple molecules.

**Importance of molecular flexibility for the resulting epitaxy.** To assess the role of molecular flexibility in epitaxial matching, we recalculated the energy landscape of BTB on the δ-layer with the molecules constrained to a planar geometry. We observe a pronounced change in the preferred adsorption positions compared to fully relaxed BTB. For planar BTB, all low-energy configurations within 40 meV per molecule of the calculated minimum are located exclusively in the empty regions of the δ-layer, consistent with the adsorption preferences of rigid HMTP molecules. This largely differs from fully relaxed BTB molecules that exhibit additional lateral positions, where less favorable molecule–substrate interactions are compensated by lower conformational energy penalties. Importantly, the preferred adsorption positions differ between the planar and relaxed configurations, demonstrating that intramolecular flexibility directly modifies the energetic preferences for lateral registry. Considering that molecular flexibility also influences optimal lattice parameters of the BTB overlayer, we conclude that molecular flexibility is an important factor that drives the epitaxy at conformationally frustrated heteromolecular interfaces.

In summary, conformational frustration arises because no single structure can simultaneously optimize intermolecular hydrogen bonding, adsorption energy on the substrate, and the intrinsic conformation in the overlayer. In general, the limiting regimes are: (I) $E_{\mathrm{int}}$ dominates the conformational cost $E_{\mathrm{conf}}$ and (II) conformational preference outweighs the adaptation to the interface. In regime I, the molecules are uniformly forced into a substrate-preferred conformation; the layer may be distorted from the optimal conformation and strained. In regime II, the molecules retain their intrinsic conformation, molecule-substrate interactions are weak, and an epitaxial relationship may not be obtained. HMTP/δ-layer is an example of regime I with dominant $E_{\mathrm{int}}$. For BTB, the conformational and interfacial energy terms are comparable; thus, the BTB/δ-layer lies in the intermediate regime (I-II). This leads to the strongest conformational frustration and the presence of an extended overlayer unit cell that epitaxially matches the substrate, with strong conformational variation among distinct molecules in the unit cell. Conversely, the BTB/Ag(111) lies closer to regime I, where the stronger interaction with the underlying metal substrate enforces planarization of the network.

## Conclusions

In conclusion, our study experimentally and computationally investigates epitaxial growth at the heteromolecular interfaces. We compare the structures of BTB molecular layers on Ag(111) and on the δ-layer. The characteristic binding motifs are principally retained on both surfaces, but the overlayer lattice parameters, orientations and conformations of individual BTB molecules change. Our computations provide a two-level description of molecular epitaxy. At the coarse structural level, intermolecular interactions define the molecular bonding motifs and the overall network structure independent of the substrate. At the fine epitaxial level, interface corrugation, molecular conformation, and strain influence the orientation registry and lattice parameters of the commensurate overlayer. In particular, the competition between molecule-substrate interactions and intramolecular conformational energy determines the relative stability of different lateral registries. Conformational frustration arises from this competition leading to distinct compromise geometries at different sites within the epitaxial unit cell. Constraining BTB to a planar geometry changes the preferred lateral adsorption positions compared to fully relaxed BTB, demonstrating that molecular flexibility directly influences the energetics of epitaxial registry. The behavior of the less flexible HMTP overlayer provides a complementary comparison. We conclude that conformational frustration becomes relevant when: (i) the molecular conformation favored by the substrate differs from the intrinsically preferred conformation, and (ii), the resulting variations in conformational and molecule-substrate interaction energies are comparable across different adsorption sites. In our system, this competition produces periodic, site-dependent conformational variations within a chemically uniform molecular overlayer. However, the same energetic competition can arise at other molecular interfaces. Our findings thus identify conformational frustration as a general mechanism through which intramolecular flexibility reshapes the energy landscape that drives the epitaxial matching.

## Methods

All experiments were conducted in an ultrahigh vacuum (UHV) system at the CEITEC Nano Research Infrastructure. The system consists of multiple UHV chambers interconnected by a UHV transfer line that enables sample transfer while maintaining a base pressure of $2\times10^{-10}$ mbar.

**Sample Preparation.** Ag(111) single-crystal substrates were prepared by repeated cycles of $Ar^+$ sputtering and annealing at 520 °C, followed by gradual cooling (< 1 °C/s) to room temperature in the preparation chamber with a base pressure of $2\times10^{-10}$ mbar. Subsequently, BTB molecules were deposited in an adjacent "deposition chamber" using a Standard Effusion Cell (WEZ40, MBE Komponenten) with a resistively heated alumina crucible held at 210 °C while the sample was maintained at room temperature. A complete BTB monolayer was achieved after 20 minutes of deposition. BTB was purchased from Merck (97% purity) and used after thorough degassing in UHV. We define 1 ML of BTB as full coverage of surface by δ-layer. Taking the areas per single molecule in distinct phases, full coverage by stripe phase occurs at 0.87 or 0.90 ML on Ag(111) and δ-layer, respectively, and by honeycomb phase at 0.35 or 0.36 ML on Ag(111) and δ-layer, respectively.

**Scanning Tunneling Microscopy (STM)** was performed using a commercial system, the Aarhus (Specs), equipped with a KolibriSensor and maintained a base pressure of $1\times10^{-10}$ mbar. STM measurements were performed at room temperature in constant current mode. The sample bias voltage ranged from 0.4 to 1.5 V, while the tunneling current was set at 50 pA. The image drift distortions were corrected by transforming the image to match the unit cell determined in STM with the one obtained from diffraction in LEEM. The STM images were rotated to align the horizontal and vertical directions with the main Ag(111) crystallographic substrate directions.

**Low-Energy Electron Microscopy/Diffraction (LEEM/LEED)** experiments were conducted using a Specs FE-LEEM P90 instrument, which maintained a base pressure of $2\times10^{-10}$ mbar. Bright-field images were measured by capturing electrons collected from the central (0,0) diffracted beam. Diffraction patterns were acquired from a surface area of 15×10 $\mu m^2$, and microdiffraction was measured using a 185 nm (diameter) e-beam spot size on the sample. The diffraction patterns were analyzed using ProLEED Studio,[65] employing a local congruence approach.

**Synchrotron Radiation Photoelectron Spectroscopy** analysis was conducted at the Materials Science Beamline of the Elettra synchrotron light source in Trieste according to the experimental protocols reported previously [10,64]. In brief, to measure the O 1s spectra, we used an excitation energy of 670 eV, a medium-area lens mode, a 10 eV pass energy, and integrated 25 sweeps with a 0.1 s dwell time and a 0.05 eV energy step. The peak positions were corrected with respect to the measured Fermi edge of the Ag substrate, and the intensity was normalized to the photo-current of a gold mesh placed in the beamline. Subsequently, the photoelectron spectra were fitted using Voigt profiles and a Shirley background.

**Computational details.** All density functional theory calculations are carried out using the Vienna *ab initio* simulation package (VASP).[66] For silver, carbon, hydrogen, and oxygen, 11 valence electrons, 4 valence electrons, 1 valence electron, and 6 valence electrons, respectively, are described with a plane-wave basis set with an energy cut-off set to 450 eV, while all core electrons are treated by the projector augmented wave method (PAW).[67] The exchange-correlation energy is described by the PBE-D3 functional[68,69] and the Brillouin zone is sampled using a single Γ point. Geometry optimization is stopped when all residual forces acting on the atoms are below 0.02 eV/Å. For calculations featuring the BTB overlayer, we remove the Ag(111) substrate completely, fix the atomic positions of the δ-layer, and saturate the deprotonated carboxylic groups of the δ-layer with hydrogen atoms placed 1 Å below the corresponding anchored oxygen atoms. This approximation is reasonable because of the weak electronic coupling between the Ag(111) surface and the BTB overlayer. The validity of removing the Ag(111) substrate and fixing the δ-layer is demonstrated in Supplementary Information, Section 17. Since the experimental supercells of the BTB overlayers are still too large for DFT calculations, we restrict our calculations to the stripe phase adsorbed on the fixed δ-layer using an alternative supercell defined by the epitaxy matrix $\begin{pmatrix} 4 & -1 \\ 0 & 1 \end{pmatrix}$ with respect to the primitive cell of the BTB overlayer, and $\begin{pmatrix} 4 & -1 \\ 1 & 2 \end{pmatrix}$with respect to the unit cell of the δ-layer, i.e., $\begin{pmatrix} 5 & 1 \\ -1 & 4 \end{pmatrix}$. This supercell induces only a minor compressive strain of ~1% along the $b$ in-plane direction of the BTB overlayer and ~1% compressive angular strain. All calculations include dipole corrections to the potential and energy.

AUTHOR INFORMATION

**Corresponding Author**

* E-mail: cechal@fme.vutbr.cz (J. Č.)

**Author Contributions**

T.K. prepared the samples, performed laboratory (LEEM, STM, and XPS) measurements on Ag(111) and evaluated STM data. P.P. together with V.S., evaluated LEEM/LEED data. V.S., P.P., T.S., and J.Č. performed the synchrotron radiation experiments. J.Č. evaluated the XPS and synchrotron data. J.P. performed the DFT calculations and evaluated the results. J.Č. coordinated the research, interpreted the results, and wrote the manuscript. All authors contributed to the manuscript draft and discussed the results.

COMPETING INTERESTS

The authors declare no competing interests.

ACKNOWLEDGMENT

This research has been supported by GAČR, project No. 23-08001S. We acknowledge CzechNanoLab Research Infrastructure (LM2023051) and e-INFRA CZ (ID:90254) supported by MEYS CR. J.P. was supported by the ERA Fellowship, Project No.: 101130765. The authors acknowledge the CERIC-ERIC Consortium for access to experimental facilities and financial support.

REFERENCES

(1) Barth, J. V.; Costantini, G.; Kern, K. Engineering Atomic and Molecular Nanostructures at Surfaces. *Nature* **2005**, *437* (7059), 671–679. https://doi.org/10.1038/nature04166.

(2) Elemans, J. A. A. W.; Lei, S.; De Feyter, S. Molecular and Supramolecular Networks on Surfaces : From Two-Dimensional Crystal Engineering to Reactivity. *Angew. Chem. Int. Ed.* **2009**, *48*, 7298–7332. https://doi.org/10.1002/anie.200806339.

(3) Goronzy, D. P.; Ebrahimi, M.; Rosei, F.; Arramel; Fang, Y.; De Feyter, S.; Tait, S. L.; Wang, C.; Beton, P. H.; Wee, A. T. S.; Weiss, P. S.; Perepichka, D. F. Supramolecular Assemblies on Surfaces: Nanopatterning, Functionality, and Reactivity. *ACS Nano* **2018**, *12*, 7445–7481.

(4) Wakayama, Y. On-Surface Molecular Nanoarchitectonics : From Self-Assembly to Directed Assembly. *Jpn. J. Appl. Phys.* **2016**, *55*, 1102AA. https://doi.org/10.7567/JJAP.55.1102AA.

(5) Dong, L.; Gao, Z. A.; Lin, N. Self-Assembly of Metal–Organic Coordination Structures on Surfaces. *Prog. Surf. Sci.* **2016**, *91*, 101–135. https://doi.org/10.1016/j.progsurf.2016.08.001 0079-6816.

(6) Bouju, X.; Mattioli, C.; Franc, G.; Pujol, A.; Gourdon, A. Bicomponent Supramolecular Architectures at the Vacuum−Solid Interface. *Chem. Rev.* **2017**, *117*, 1407–1444. https://doi.org/10.1021/acs.chemrev.6b00389.

(7) Otero, R.; Vázquez de Parga, A. L.; Gallego, J. M. Electronic, Structural and Chemical Effects of Charge-Transfer at Organic/Inorganic Interfaces. *Surf. Sci. Rep.* **2017**, *72*, 105–145. https://doi.org/10.1016/j.surfrep.2017.03.001.

(8) Urgel, J. I.; Écija, D.; Lyu, G.; Zhang, R.; Palma, C.-A.; Auwärter, W.; Lin, N.; Barth, J. V. Quasicrystallinity Expressed in Two-Dimensional Coordination Networks. *Nat.*

*Chem.* **2016**, *8* (7), 657–662. https://doi.org/10.1038/nchem.2507.

(9) Ernst, K.-H. Stereochemistry of 2D Molecular Crystallization. *Chim. Int. J. Chem.* **2014**, *68* (1), 49–53. https://doi.org/10.2533/chimia.2014.49.

(10) Stará, V.; Procházka, P.; Planer, J.; Shahsavar, A.; Makoveev, A. O.; Skála, T.; Blatnik, M.; Čechal, J. Tunable Energy-Level Alignment in Multilayers of Carboxylic Acids on Silver. *Phys. Rev. Appl.* **2022**, *18* (4), 044048. https://doi.org/10.1103/PhysRevApplied.18.044048.

(11) Pan, M.; Zhang, X.; Zhou, Y.; Wang, P.; Bian, Q.; Liu, H.; Wang, X.; Li, X.; Chen, A.; Lei, X.; Li, S.; Cheng, Z.; Shao, Z.; Ding, H.; Gao, J.; Li, F.; Liu, F. Growth of Mesoscale Ordered Two-Dimensional Hydrogen-Bond Organic Framework with the Observation of Flat Band. *Phys. Rev. Lett.* **2023**, *130* (3), 036203. https://doi.org/10.1103/PhysRevLett.130.036203.

(12) Lobo-Checa, J.; Hernández-López, L.; Otrokov, M. M.; Piquero-Zulaica, I.; Candia, A. E.; Gargiani, P.; Serrate, D.; Delgado, F.; Valvidares, M.; Cerdá, J.; Arnau, A.; Bartolomé, F. Ferromagnetism on an Atom-Thick & Extended 2D Metal-Organic Coordination Network. *Nat. Commun.* **2024**, *15* (1), 1858. https://doi.org/10.1038/s41467-024-46115-z.

(13) Parreiras, S. O.; Martín-Fuentes, C.; Moreno, D.; Mathialagan, S. K.; Biswas, K.; Muñiz-Cano, B.; Lauwaet, K.; Valvidares, M.; Valbuena, M. A.; Urgel, J. I.; Gargiani, P.; Camarero, J.; Miranda, R.; Martínez, J. I.; Gallego, J. M.; Écija, D. 2D Co-Directed Metal–Organic Networks Featuring Strong Antiferromagnetism and Perpendicular Anisotropy. *Small* **2024**, *20* (22), 2309555. https://doi.org/10.1002/smll.202309555.

(14) Frezza, F.; Matěj, A.; Sánchez-Grande, A.; Carrera, M.; Mutombo, P.; Kumar, M.; Curiel, D.; Jelínek, P. On-Surface Synthesis of a Radical 2D Supramolecular Organic Framework. *J. Am. Chem. Soc.* **2024**, *146*, 3531–3538.

https://doi.org/10.1021/jacs.3c13702.

(15) Wang, D.; Wang, Z.; Wu, S.; Arramel; Yin, X.; Tang, C. S.; Feng, Y. P.; Wu, J.; Wee, A. T. S. Realizing Two-Dimensional Supramolecular Arrays of a Spin Molecule via Halogen Bonding. *ACS Nanosci. Au* **2022**, *2* (4), 333–340. https://doi.org/10.1021/acsnanoscienceau.2c00005.

(16) Piquero-Zulaica, I.; Lobo-Checa, J.; El-Fattah, Z. M. A.; Ortega, J. E.; Klappenberger, F.; Auwärter, W.; Barth, J. V. Engineering Quantum States and Electronic Landscapes through Surface Molecular Nanoarchitectures. *Rev. Mod. Phys.* **2022**, *94* (4), 045008. https://doi.org/10.1103/RevModPhys.94.045008.

(17) Kawai, S.; Kher-Elden, M. A.; Sadeghi, A.; Abd El-Fattah, Z. M.; Sun, K.; Izumi, S.; Minakata, S.; Takeda, Y.; Lobo-Checa, J. Near Fermi Superatom State Stabilized by Surface State Resonances in a Multiporous Molecular Network. *Nano Lett.* **2021**, *21* (15), 6456–6462. https://doi.org/10.1021/acs.nanolett.1c01200.

(18) Hu, T.; Zhong, W.; Zhang, T.; Wang, W.; Wang, Z. F. Identifying Topological Corner States in Two-Dimensional Metal-Organic Frameworks. *Nat. Commun.* **2023**, *14* (1), 7092. https://doi.org/10.1038/s41467-023-42884-1.

(19) Coates, S.; Smerdon, J. A.; McGrath, R.; Sharma, H. R. A Molecular Overlayer with the Fibonacci Square Grid Structure. *Nat. Commun.* **2018**, *9* (1), 3435. https://doi.org/10.1038/s41467-018-05950-7.

(20) Graham, M. J.; Zadrozny, J. M.; Fataftah, M. S.; Freedman, D. E. Forging Solid-State Qubit Design Principles in a Molecular Furnace. *Chem. Mater.* **2017**, *29*, 1885–1897. https://doi.org/10.1021/acs.chemmater.6b05433.

(21) Jakub, Z.; Shahsavar, A.; Planer, J.; Hrůza, D.; Herich, O.; Procházka, P.; Čechal, J. How the Support Defines Properties of 2D Metal–Organic Frameworks: Fe-TCNQ on Graphene versus Au(111). *J. Am. Chem. Soc.* **2024**, *146* (5), 3471–3482.

https://doi.org/10.1021/jacs.3c13212.

(22) Jakub, Z.; Planer, J.; Hrůza, D.; Shahsavar, A.; Pavelec, J.; Čechal, J. Identical Fe–N 4 Sites with Different Reactivity: Elucidating the Effect of Support Curvature. *ACS Appl. Mater. Interfaces* **2025**, *17* (6), 10136–10144. https://doi.org/10.1021/acsami.4c19913.

(23) Wurster, B.; Grumelli, D.; Hötger, D.; Gutzler, R.; Kern, K. Driving the Oxygen Evolution Reaction by Nonlinear Cooperativity in Bimetallic Coordination Catalysts. *J. Am. Chem. Soc.* **2016**, *138*, 3623–3626. https://doi.org/10.1021/jacs.5b10484.

(24) Gutzler, R.; Stepanow, S.; Grumelli, D.; Lingenfelder, M.; Kern, K. Mimicking Enzymatic Active Sites on Surfaces for Energy Conversion Chemistry. *Acc. Chem. Res.* **2015**, *48*, 2132–2139. https://doi.org/10.1021/acs.accounts.5b00172.

(25) Li, C.; Zhang, X.; Li, N.; Wang, Y.; Yang, J.; Gu, G.; Zhang, Y.; Hou, S.; Peng, L.; Wu, K.; Nieckarz, D.; Szabelski, P.; Tang, H.; Wang, Y. Construction of Sierpiński Triangles up to the Fifth Order. *J. Am. Chem. Soc.* **2017**, *139* (39), 13749–13753. https://doi.org/10.1021/jacs.7b05720.

(26) Impéror-Clerc, M.; Jagannathan, A.; Kalugin, P.; Sadoc, J.-F. Square-Triangle Tilings: An Infinite Playground for Soft Matter. *Soft Matter* **2021**, *17* (42), 9560–9575. https://doi.org/10.1039/D1SM01242H.

(27) Ecija, D.; Urgel, J. I.; Papageorgiou, A. C.; Joshi, S.; Auwarter, W.; Seitsonen, A. P.; Klyatskaya, S.; Ruben, M.; Fischer, S.; Vijayaraghavan, S.; Reichert, J.; Barth, J. V. Five-Vertex Archimedean Surface Tessellation by Lanthanide-Directed Molecular Self-Assembly. *Proc. Natl. Acad. Sci.* **2013**, *110* (17), 6678–6681. https://doi.org/10.1073/pnas.1222713110.

(28) Kormoš, L.; Procházka, P.; Makoveev, A. O.; Čechal, J. Complex K-Uniform Tilings by a Simple Bitopic Precursor Self-Assembled on Ag(001) Surface. *Nat. Commun.* **2020**, *11*, 1856. https://doi.org/10.1038/s41467-020-15727-6.

(29) Blunt, M. O.; Russell, J. C.; Gimenez-Lopez, M. d. C.; Garrahan, J. P.; Lin, X.; Schroder, M.; Champness, N. R.; Beton, P. H. Random Tiling and Topological Defects in a Two-Dimensional Molecular Network. *Science (80-. ).* **2008**, *322* (5904), 1077–1081. https://doi.org/10.1126/science.1163338.

(30) Stannard, A.; Russell, J. C.; Blunt, M. O.; Salesiotis, C.; Giménez-López, M. D. C.; Taleb, N.; Schröder, M.; Champness, N. R.; Garrahan, J. P.; Beton, P. H. Broken Symmetry and the Variation of Critical Properties in the Phase Behaviour of Supramolecular Rhombus Tilings. *Nat. Chem.* **2012**, *4* (2), 112–117. https://doi.org/10.1038/nchem.1199.

(31) Xu, Y.; Duan, J.-J.; Yi, Z.-Y.; Zhang, K.-X.; Chen, T.; Wang, D. Chirality of Molecular Nanostructures on Surfaces via Molecular Assembly and Reaction: Manifestation and Control. *Surf. Sci. Rep.* **2021**, *76*, 100531. https://doi.org/10.1016/j.surfrep.2021.100531.

(32) Ding, P.; Wang, S.; Mattioli, C.; Li, Z.; Shi, G.; Sun, Y.; Gourdon, A.; Kantorovich, L.; Besenbacher, F.; Rosei, F.; Yu, M. Extending On-Surface Synthesis from 2D to 3D by Cycloaddition with C60. *Nat. Commun.* **2023**, *14* (1), 6075. https://doi.org/10.1038/s41467-023-41913-3.

(33) Korolkov, V. V.; Baldoni, M.; Watanabe, K.; Taniguchi, T.; Besley, E.; Beton, P. H. Supramolecular Heterostructures Formed by Sequential Epitaxial Deposition of Two-Dimensional Hydrogen-Bonded Arrays. *Nat. Chem.* **2017**, *9* (12), 1191–1197. https://doi.org/10.1038/nchem.2824.

(34) Cojal González, J. D.; Iyoda, M.; Rabe, J. P. Templated Bilayer Self-Assembly of Fully Conjugated π-Expanded Macrocyclic Oligothiophenes Complexed with Fullerenes. *Nat. Commun.* **2017**, *8* (1), 14717. https://doi.org/10.1038/ncomms14717.

(35) Wang, Z.; Qian, K.; Öner, M. A.; Deimel, P. S.; Wang, Y.; Zhang, S.; Zhang, X.;

Gupta, V.; Li, J.; Gao, H.-J.; Duncan, D. A.; Barth, J. V.; Lin, X.; Allegretti, F.; Du, S.; Palma, C.-A. Layer-by-Layer Epitaxy of Porphyrin−Ligand Fe(II)-Fe(III) Nanoarchitectures for Advanced Metal–Organic Framework Growth. *ACS Appl. Nano Mater.* **2020**, *3* (12), 11752–11759. https://doi.org/10.1021/acsanm.0c02237.

(36) Zhan, G.; Koek, B.; Yuan, Y.; Liu, Y.; Mishra, V.; Lenzi, V.; Strutyński, K.; Li, C.; Zhang, R.; Zhou, X.; Choi, H. S.; Cai, Z.-F.; Almarza, J.; Mali, K. S.; Mateo-Alonso, A.; Franco, M. M.; Zhu, Y.; De Feyter, S.; Loh, K. P. Moiré Two-Dimensional Covalent Organic Framework Superlattices. *Nat. Chem.* **2025**. https://doi.org/10.1038/s41557-025-01748-5.

(37) Li, J.; Wieghold, S.; Öner, M. A.; Simon, P.; Hauf, M. V.; Margapoti, E.; Garrido, J. A.; Esch, F.; Palma, C.-A.; Barth, J. V. Three-Dimensional Bicomponent Supramolecular Nanoporous Self-Assembly on a Hybrid All-Carbon Atomically Flat and Transparent Platform. *Nano Lett.* **2014**, *14* (8), 4486–4492. https://doi.org/10.1021/nl501452s.

(38) Cao, H.; De Feyter, S. Amplification of Chirality in Surface-Confined Supramolecular Bilayers. *Nat. Commun.* **2018**, *9* (1), 3416. https://doi.org/10.1038/s41467-018-05962-3.

(39) Marocchi, S.; Ferriani, P.; Caffrey, N. M.; Manghi, F.; Heinze, S.; Bellini, V. Graphene-Mediated Exchange Coupling between a Molecular Spin and Magnetic Substrates. *Phys. Rev. B* **2013**, *88*, 144407. https://doi.org/10.1103/PhysRevB.88.144407.

(40) Cojocariu, I.; Jugovac, M.; Sarwar, S.; Rawson, J.; Sanz, S.; Kögerler, P.; Feyer, V.; Schneider, C. M. Semiconductor Halogenation in Molecular Highly-Oriented Layered p–n (n–p) Junctions. *Adv. Funct. Mater.* **2022**, *32* (51), 2208507. https://doi.org/10.1002/adfm.202208507.

(41) Sato, Y.; De Feyter, S.; Tahara, K. Formation of Supramolecular Heterostacks at the Liquid–Solid Interface: Impact of Symmetry Mismatching on Structural Growth. *Langmuir* **2023**, *39* (47), 16825–16832. https://doi.org/10.1021/acs.langmuir.3c02327.

(42) Fang, Y.; Cibian, M.; Hanan, G. S.; Perepichka, D. F.; De Feyter, S.; Cuccia, L. A.; Ivasenko, O. Alkyl Chain Length Effects on Double-Deck Assembly at a Liquid/Solid Interface. *Nanoscale* **2018**, *10* (31), 14993–15002. https://doi.org/10.1039/C8NR04220A.

(43) Procházka, P.; Stará, V.; Planer, J.; Skála, T.; Čechal, J. Structure of Epitaxial Heteromolecular Bilayers Probed by Low-Energy Electron Microscopy. *Surfaces and Interfaces* **2025**, *62*, 106251. https://doi.org/10.1016/j.surfin.2025.106251.

(44) Meissner, M.; Sojka, F.; Matthes, L.; Bechstedt, F.; Feng, X.; Müllen, K.; Mannsfeld, S. C. B.; Forker, R.; Fritz, T. Flexible 2D Crystals of Polycyclic Aromatics Stabilized by Static Distortion Waves. *ACS Nano* **2016**, *10* (7), 6474–6483. https://doi.org/10.1021/acsnano.6b00935.

(45) Bauer, E. Surface Microscopy with Low Energy Electrons: LEEM. *J. Electron Spectros. Relat. Phenomena* **2020**, *241*, 146806. https://doi.org/10.1016/j.elspec.2018.11.005.

(46) Tromp, R. M. Low-Energy Electron Microscopy. *IBM J. Res. Dev.* **2000**, *44*, 503–516. https://doi.org/10.1147/rd.444.0503.

(47) Čechal, J.; Procházka, P. Low-Energy Electron Microscopy as a Tool for Analysis of Self-Assembled Molecular Layers on Surfaces. *J. Phys. Condens. Matter* **2025**, *37*, 293003. https://doi.org/10.1088/1361-648X/ade946.

(48) Čechal, J. Essential Principles and Practices in X-Ray Photoelectron Spectroscopy. *J. Vac. Sci. Technol. A* **2026**, *44* (4). https://doi.org/10.1116/6.0005469.

(49) Krajňák, T.; Stará, V.; Procházka, P.; Planer, J.; Skála, T.; Blatnik, M.; Čechal, J.

Robust Dipolar Layers between Organic Semiconductors and Silver for Energy-Level Alignment. *ACS Appl. Mater. Interfaces* **2024**, *16*, 18099–18111. https://doi.org/10.1021/acsami.3c18697.

(50) Hieulle, J.; Mura, M.; Silly, F. Assessment of Intermolecular Hydrogen Bonds in the Compact Self-Assemblies of 1,3,5-Tris(4-Carboxyphenyl)Benzene Molecules on Au(111). *Langmuir* **2024**, *40*, 26455–26462. https://doi.org/10.1021/acs.langmuir.4c03099.

(51) Silly, F. Two-Dimensional 1,3,5-Tris(4-Carboxyphenyl)Benzene Self-Assembly at the 1-Phenyloctane/Graphite Interface Revisited. *J. Phys. Chem. C* **2012**, *116* (18), 10029–10032. https://doi.org/10.1021/jp300678m.

(52) Čechal, J.; Stará, V.; Kurowská, A.; Blatnik, M.; Pravec, D.; Hrubá, D.; Kunc, J.; Drašar, Č.; Planer, J.; Procházka, P. Structural, Electronic, and Chemical Manifestations of Weak Molecule–Substrate Coupling on Bi2Se3 Compared with Ag, Au, and Graphene. *Appl. Surf. Sci. Adv.* **2026**, *32*, 100927. https://doi.org/10.1016/j.apsadv.2025.100927.

(53) Ruben, M.; Payer, D.; Landa, A.; Comisso, A.; Gattinoni, C.; Lin, N.; Collin, J.-P.; Sauvage, J.-P.; De Vita, A.; Kern, K. 2D Supramolecular Assemblies of Benzene-1,3,5-Triyl-Tribenzoic Acid: Temperature-Induced Phase Transformations and Hierarchical Organization with Macrocyclic Molecules. *J. Am. Chem. Soc.* **2006**, *128* (49), 15644–15651. https://doi.org/10.1021/ja063601k.

(54) Makoveev, A. O.; Procházka, P.; Blatnik, M.; Kormoš, L.; Skála, T.; Čechal, J. Role of Phase Stabilization and Surface Orientation in 4,4′-Biphenyl-Dicarboxylic Acid Self-Assembly and Transformation on Silver Substrates. *J. Phys. Chem. C* **2022**, *126*, 9989–9997. https://doi.org/10.1021/acs.jpcc.2c02538.

(55) Quiroga Argañaraz, B.; Cristina, L. J.; Rodríguez, L. M.; Cossaro, A.; Verdini, A.;

Floreano, L.; Fuhr, J. D.; Gayone, J. E.; Ascolani, H. Ubiquitous Deprotonation of Terephthalic Acid in the Self-Assembled Phases on Cu(100). *Phys. Chem. Chem. Phys.* **2018**, *20*, 4329–4339. https://doi.org/10.1039/c7cp06612k.

(56) Carter-Fenk, K.; Herbert, J. M. Reinterpreting π-Stacking. *Phys. Chem. Chem. Phys.* **2020**, *22*, 24870–24886. https://doi.org/10.1039/d0cp05039c.

(57) Bechgaard, K.; Parker, V. D. Mono-, Di-, and Trications of Hexamethoxytriphenylene. Novel Anodic Trimerization. *J. Am. Chem. Soc.* **1972**, *94* (13), 4749–4750. https://doi.org/10.1021/ja00768a063.

(58) Andresen, T. L.; Krebs, F. C.; Thorup, N.; Bechgaard, K. Crystal Structures of 2,3,6,7,10,11-Oxytriphenylenes. Implications for Columnar Discotic Mesophases. *Chem. Mater.* **2000**, *12* (8), 2428–2433. https://doi.org/10.1021/cm0003020.

(59) Bedwani, S.; Wegner, D.; Crommie, M. F.; Rochefort, A. Strongly Reshaped Organic-Metal Interfaces: Tetracyanoethylene on Cu(100). *Phys. Rev. Lett.* **2008**, *101*, 216105. https://doi.org/10.1103/PhysRevLett.101.216105.

(60) Tseng, T.-C.; Urban, C.; Wang, Y.; Otero, R.; Tait, S. L.; Alcamí, M.; Écija, D.; Trelka, M.; Gallego, J. M.; Lin, N.; Konuma, M.; Starke, U.; Nefedov, A.; Langner, A.; Wöll, C.; Herranz, M. Á.; Martín, F.; Martín, N.; Kern, K.; Miranda, R. Charge-Transfer-Induced Structural Rearrangements at Both Sides of Organic/Metal Interfaces. *Nat. Chem.* **2010**, *2*, 374–379. https://doi.org/10.1038/nchem.591.

(61) Kley, C. S.; Čechal, J.; Kumagai, T.; Schramm, F.; Ruben, M.; Stepanow, S.; Kern, K. Highly Adaptable Two-Dimensional Metal-Organic Coordination Networks on Metal Surfaces. *J. Am. Chem. Soc.* **2012**, *134*, 6072–6075. https://doi.org/10.1021/ja211749b.

(62) White, T. W.; Martsinovich, N.; Troisi, A.; Costantini, G. Quantifying the "Subtle Interplay" between Intermolecular and Molecule–Substrate Interactions in Molecular Assembly on Surfaces. *J. Phys. Chem. C* **2018**, *122*, 17954–17962.

https://doi.org/10.1021/acs.jpcc.8b06797.

(63) Procházka, P.; Gosalvez, M. A.; Kormoš, L.; de la Torre, B.; Gallardo, A.; Alberdi-Rodriguez, J.; Chutora, T.; Makoveev, A. O.; Shahsavar, A.; Arnau, A.; Jelínek, P.; Čechal, J. Multiscale Analysis of Phase Transformations in Self-Assembled Layers of 4,4′-Biphenyl Dicarboxylic Acid on the Ag(001) Surface. *ACS Nano* **2020**, *14*, 7269–7279. https://doi.org/10.1021/acsnano.0c02491.

(64) Procházka, P.; Kormoš, L.; Shahsavar, A.; Stará, V.; Makoveev, A. O.; Skála, T.; Blatnik, M.; Čechal, J. Phase Transformations in a Complete Monolayer of 4,4′-Biphenyl-Dicarboxylic Acid on Ag(0 0 1). *Appl. Surf. Sci.* **2021**, *547*, 149115. https://doi.org/10.1016/j.apsusc.2021.149115.

(65) Procházka, P.; Čechal, J. ProLEED Studio : Software for Modeling Low-Energy Electron Diffraction Patterns. *J. Appl. Crystallogr.* **2024**, *57* (1), 187–193. https://doi.org/10.1107/S1600576723010312.

(66) Kresse, G.; Hafner, J. Ab Initio Molecular Dynamics for Liquid Metals. *Phys. Rev. B* **1993**, *47*, 558–561.

(67) Kresse, G.; Joubert, D. From Ultrasoft Pseudopotentials to the Projector Augmented-Wave Method. *Phys. Rev. B* **1999**, *59*, 1758–1775. https://doi.org/10.1103/PhysRevB.59.1758.

(68) Perdew, J. P.; Burke, K.; Ernzerhof, M. Generalized Gradient Approximation Made Simple. *Phys. Rev. Lett.* **1996**, *77*, 3865–3868. https://doi.org/10.1103/PhysRevLett.77.3865.

(69) Grimme, S.; Antony, J.; Ehrlich, S.; Krieg, H. A Consistent and Accurate Ab Initio Parametrization of Density Functional Dispersion Correction (DFT-D) for the 94 Elements H-Pu. *J. Chem. Phys.* **2010**, *132*, 154104. https://doi.org/10.1063/1.3382344.

SUPPORTING INFORMATION

# Conformationally frustrated epitaxy at heteromolecular interfaces

*Jakub Planer,[1] Pavel Procházka,[1] Tomáš Krajňák,[1] Veronika Stará,[1] Daniela Hrubá, [1] Zdeněk Endstrasser, [1] Tomáš Skála,[2] Jan Čechal[1,3]**

[1] CEITEC - Central European Institute of Technology, Brno University of Technology, Purkyňova 123, 612 00 Brno, Czech Republic.

[2] Department of Surface and Plasma Science, Faculty of Mathematics and Physics, Charles University, V Holešovičkách 2, 180 00 Prague 8, Czech Republic.

[3] Institute of Physical Engineering, Brno University of Technology, Technická 2896/2, 616 69 Brno, Czech Republic.

CONTENTS:

## 1. Bright and dark-field analysis of the δ-layer

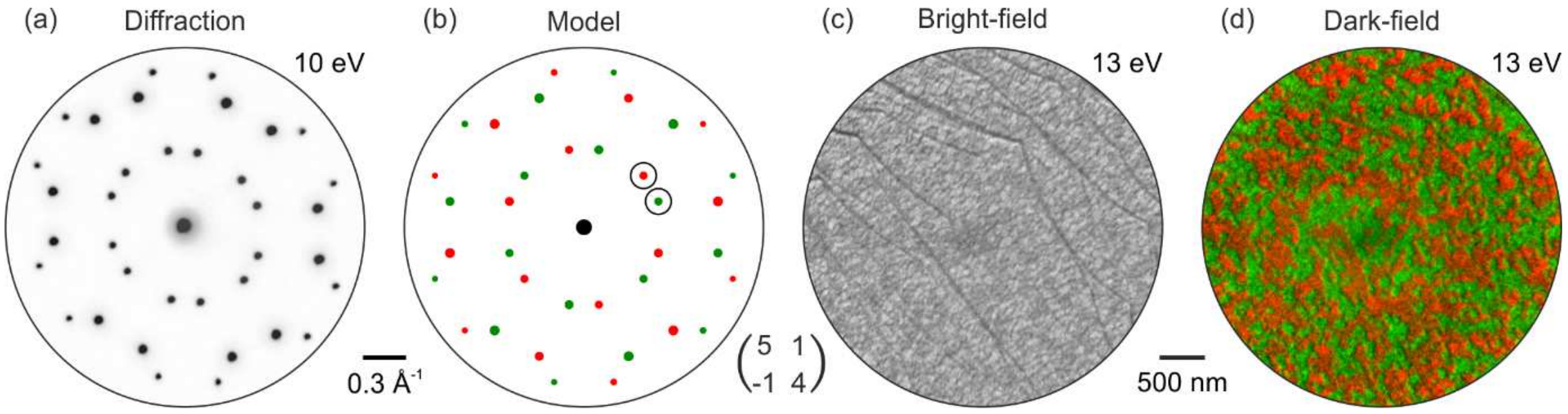

**Figure S1.** LEEM analysis of the δ-layer on Ag(111). (a) Diffraction pattern and (b) corresponding diffraction model with color-coded symmetry-equivalent domains. (c) Bright-field LEEM image of the δ-layer uniformly covering the surface. (d) Color-coded composition of dark-field LEEM images revealing the spatial distribution of the symmetry-equivalent domains in the δ-layer; the imaging was performed with an aperture positioned on the spots marked in (b).

## 2. Annealing of the BTB on the δ-layer

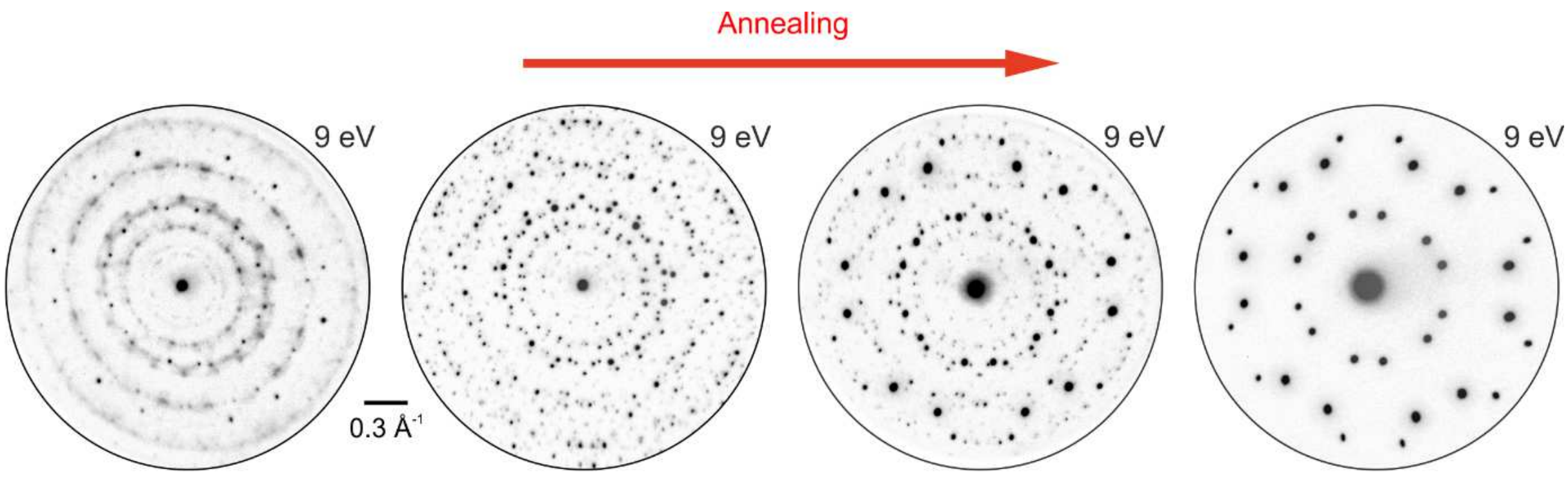

**Figure S2.** Sequence of diffraction patterns obtained during annealing of 1.0 ML of BTB deposited on the δ-layer on Ag(111). First, we observe sharpening of the diffraction pattern when all traces of honeycomb-phase diffraction disappear, and the diffraction pattern comprises only stripe-phase spots coexisting with those of the δ-layer. During annealing at 210–220 °C, BTB molecules desorb, resulting in a weakening of the associated stripe diffraction pattern. The final phase is the intact δ-BTB layer.

### 3. Single domain diffraction of the honeycomb phase on the δ-layer

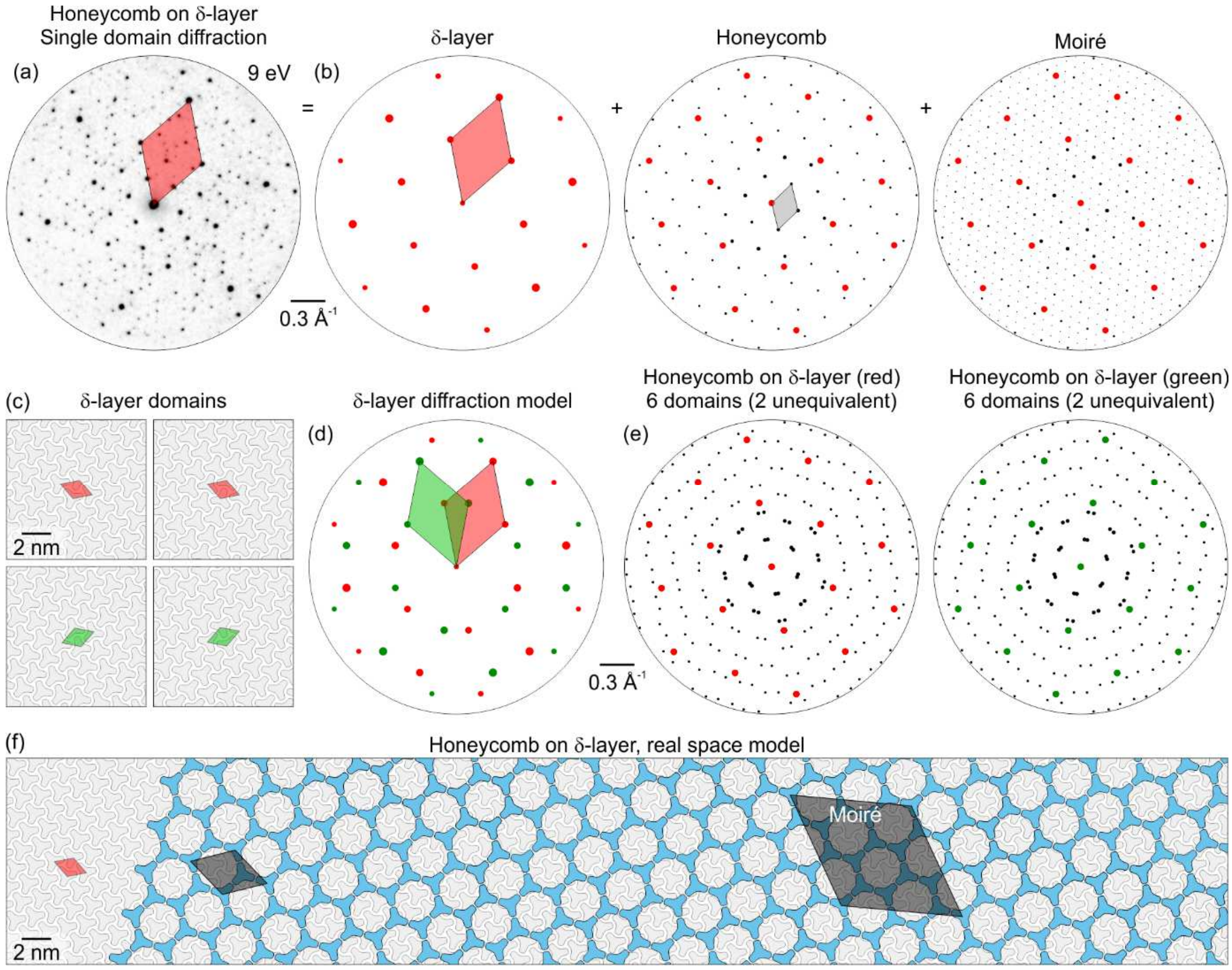

**Figure S3.** (a) Single domain diffraction pattern of honeycomb phase on δ-layer. (b) Simulated diffraction model illustrating contributions from the underlying δ-layer, honeycomb lattice, and Moiré pattern spots. (c) Four possible configurations of BTB molecules within the δ-layer. Each pair of unit cells (marked with the same color) features differently oriented BTB molecules; this gives the diffraction patterns apparent six-fold rotational symmetry while the associated molecular structure is three-fold symmetric. (d) Diffraction model of the δ-layer with color-coded domains representing symmetry-equivalent orientations. (e) Each of the two δ-layer orientations (red and green) serves as a substrate for six distinct honeycomb domains. This results from the apparent six-fold symmetry described in (c). (f) Real space structural model of the honeycomb phase on the δ-layer, corresponding to the single domain diffraction pattern shown in (a).

### 4. Comparison of unit cell sizes

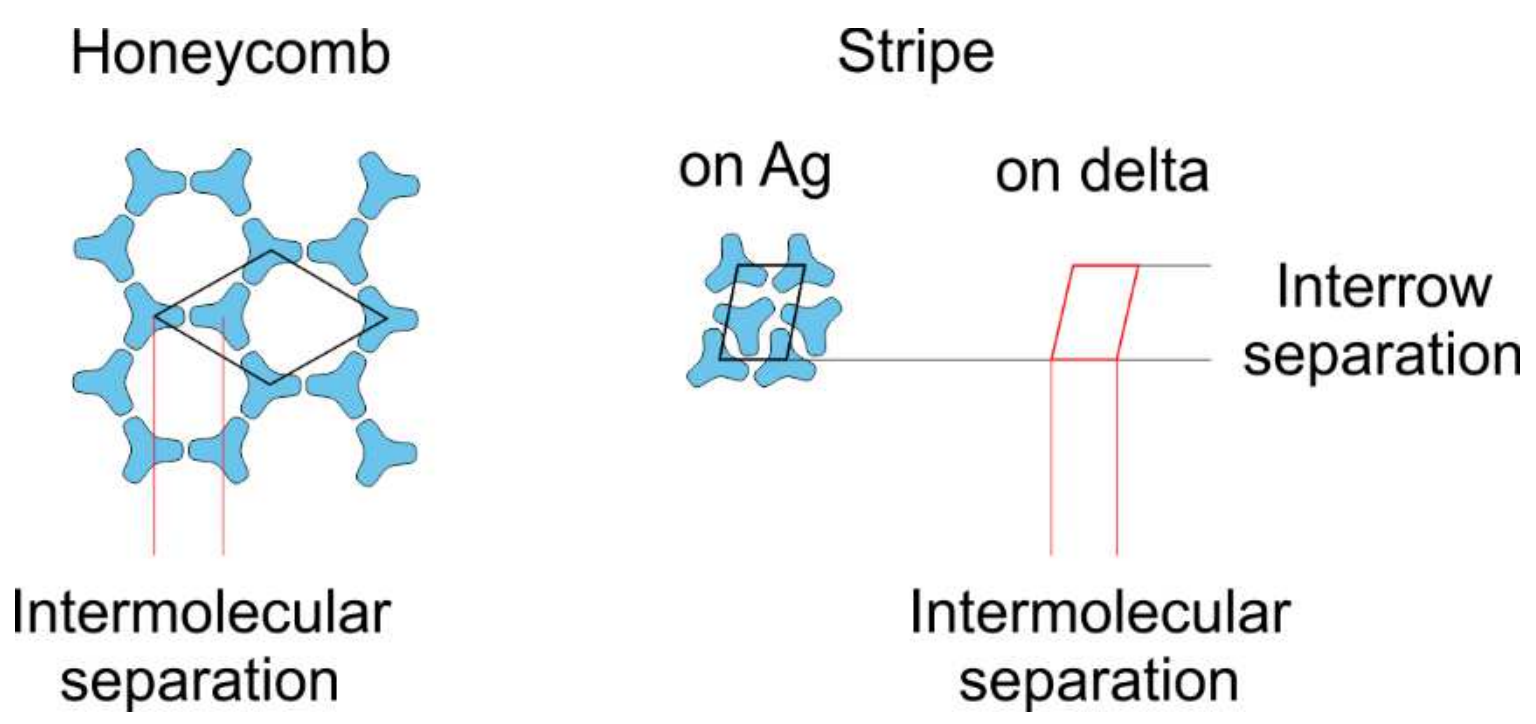

**Figure S4.** Definition of intermolecular distances discussed in the main text.

**Table S1**: Unit cell size and associated angle between unit cell vectors for honeycomb and stripe phases directly on Ag(111) and on the δ-layer obtained from the epitaxial matrices.

| Unit cell sizes (Å) | Stripe | Honeycomb |
|---|---|---|
| **on Ag(111)** | 22.57 × 15.73 × 79.01° | 31.49 × 31.49 × 120° |
| **on δ-layer** | 22.65 × 15.29 × 76.98° | 31.25 × 31.25 × 120° |

**Table S2**: Area per BTB molecule in the δ-layer, stripe, γ-stripe, and honeycomb phases on Ag(111), and in the stripe and honeycomb phases on the δ-layer obtained from the epitaxial matrices.

| Area per 1 molecule ($Å^2$) | δ-layer | Stripe | γ-stripe | Honeycomb |
|---|---|---|---|---|
| **on Ag(111)** | 152 | 174 | 152 | 429 |
| **on δ-layer** | - | 169 | - | 423 |

## 5. Building the stripe phase

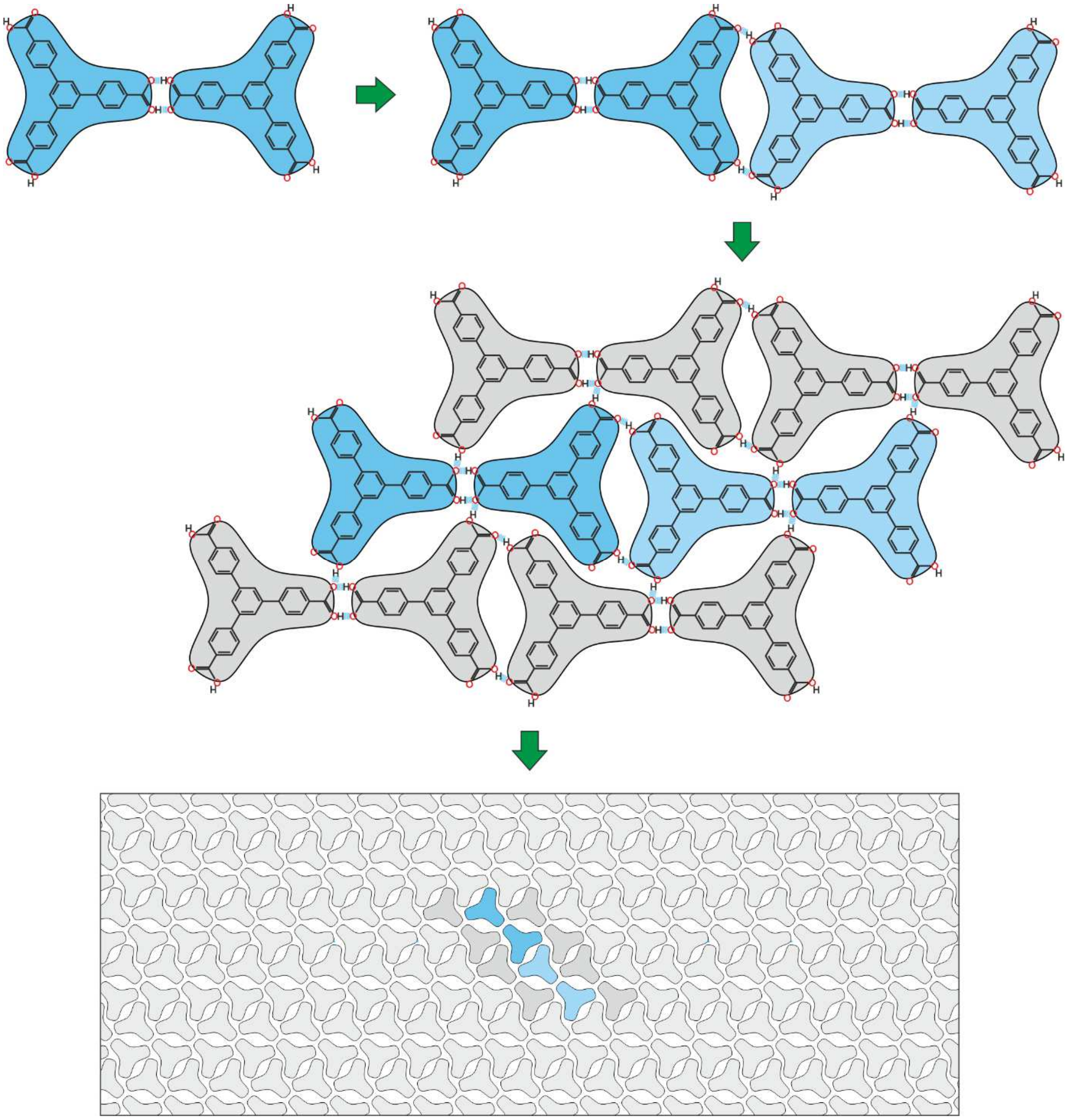

**Figure S5.** Building the stripe phase from a BTB molecular dimer.

## 6. Stripe phase on Au(111)

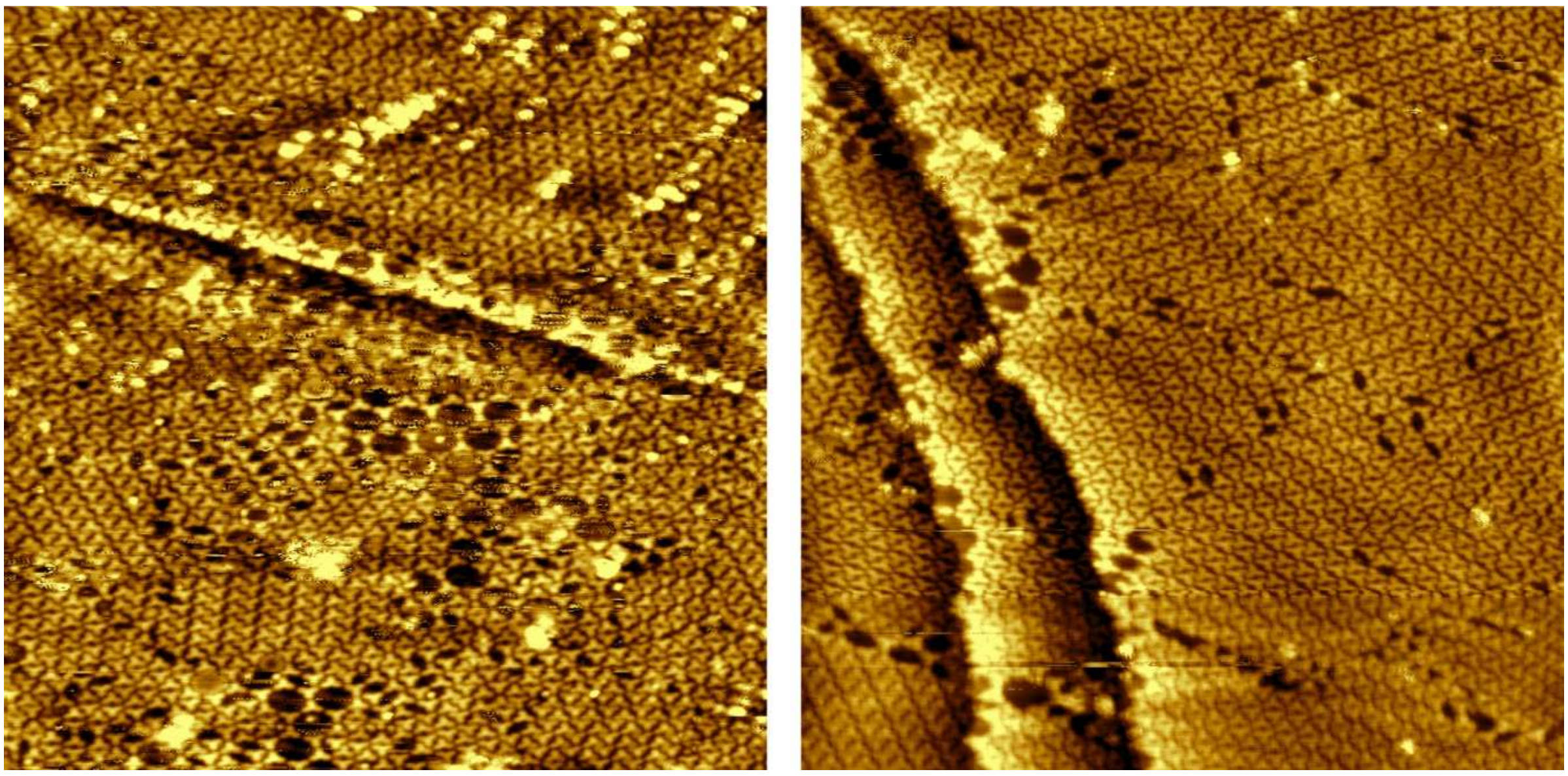

**Figure S6.** Stripe phase formed from BTB molecules on the Au(111) surface. Annealing (right image) increases the phase uniformity. On Au(111), we do not expect any deprotonation of BTB molecules. Both STM images ($50 \times 50$ nm$^2$, 1.2 V, 50 pA) are presented as acquired, without drift correction.

## 7. DFT calculations of the unsupported stripe and honeycomb phases

To assess the stability of the stripe and honeycomb phases, we have performed DFT calculations of the planar and distorted configurations using PBE-D3, PBE-TS, and optB86b functionals, as listed in Table S3. All functionals consistently predict that the distorted configurations are 250–300 meV per BTB molecule more stable than the planarized phases. While the stripe phase and honeycomb phase are energetically almost identical, with differences smaller than 36 meV per BTB molecule, the actual ground state is functional-dependent. PBE-D3 functional favors the corrugated honeycomb phase by 36 meV per BTB, PBE-TS shows that the phases are degenerate, and the optB86b functional favors the stripe phase by 14 meV per BTB.

**Table S3:** Calculated total energies of the unsupported BTB phases relative to the ground state using PBE-D3, PBE-TS, and optB86b functionals.

| $\Delta E_{tot}$ [meV / BTB] | PBE-D3 | PBE-TS | optB86b |
|---|---|---|---|
| Stripe planar | 267 | 290 | 279 |
| Stripe corrugated | 36 | 1 | 0 |
| Honeycomb planar | 246 | 301 | 287 |
| Honeycomb corrugated | 0 | 0 | 14 |

Figure S7 shows the calculated binding energies for the neighboring BTB molecules in the honeycomb and in the stripe configuration. The honeycomb structure shows a single binding motif, in which the carboxylic groups of neighboring BTB molecules point toward each other and form two hydrogen bonds. This interaction, calculated with the PBE-D3 functional, yields a binding energy of –1.14 eV. Such interaction with similar binding energy is also present in the stripe phase, but only in one of three carboxylic groups of a BTB molecule. The remaining two carboxylic groups form a single hydrogen bond with the neighboring BTB molecules, which results in weaker interactions, as shown in Figure S7b. However, there is an additional 'side' interaction between the carboxyl oxygen and the hydrogen atoms around the center of the BTB molecule (see red arrow in Figure S7b), which counterbalances the weaker interactions between the carboxylic groups in the stripe phase.

In summary, the honeycomb phase is composed of fewer interactions with strong binding, whereas the stripe phase shows more interactions with weaker binding.

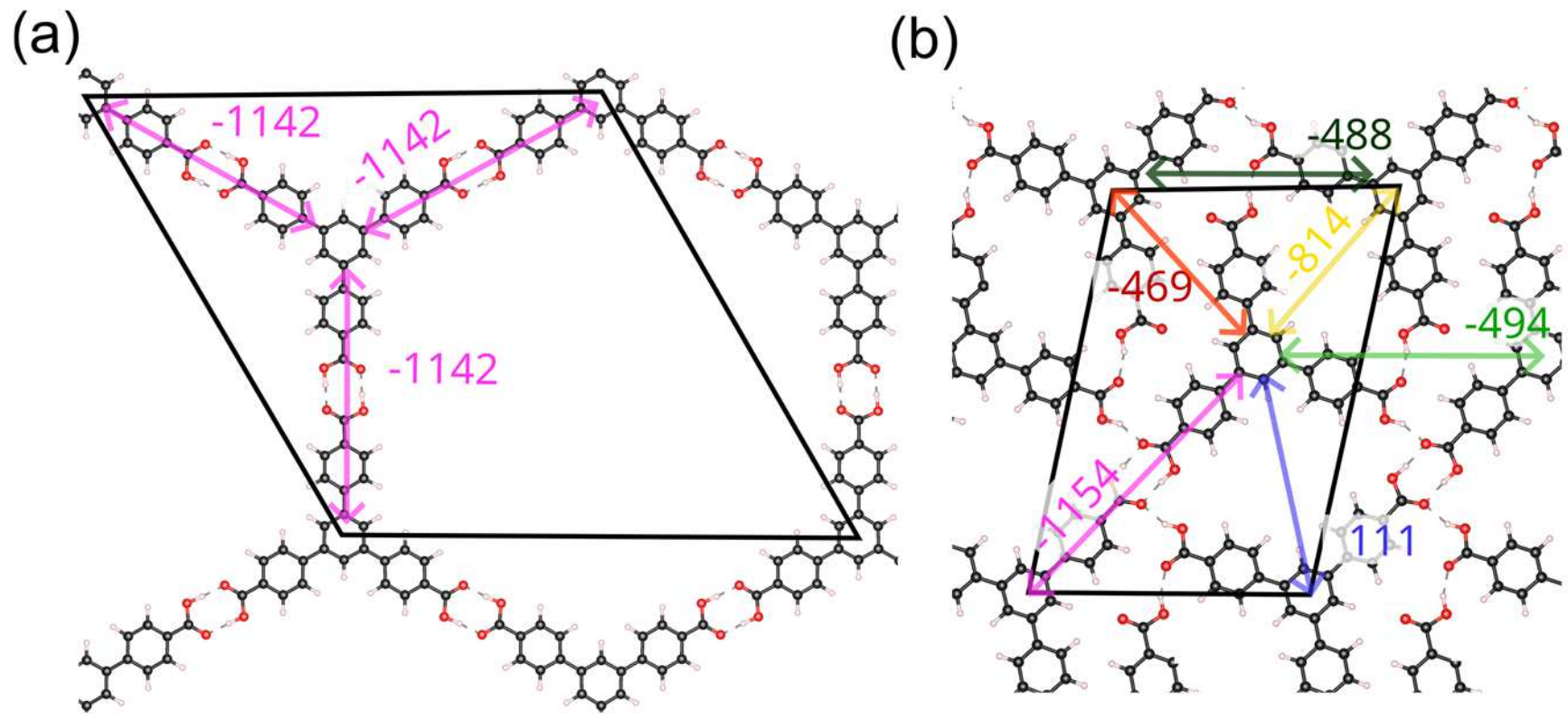

**Figure S7:** Calculated binding energies (in meV) for the neighboring BTB molecules in the honeycomb phase (a), and in the stripe phase (b).

## 8. Periodicity of the stripe phase on the δ-layer

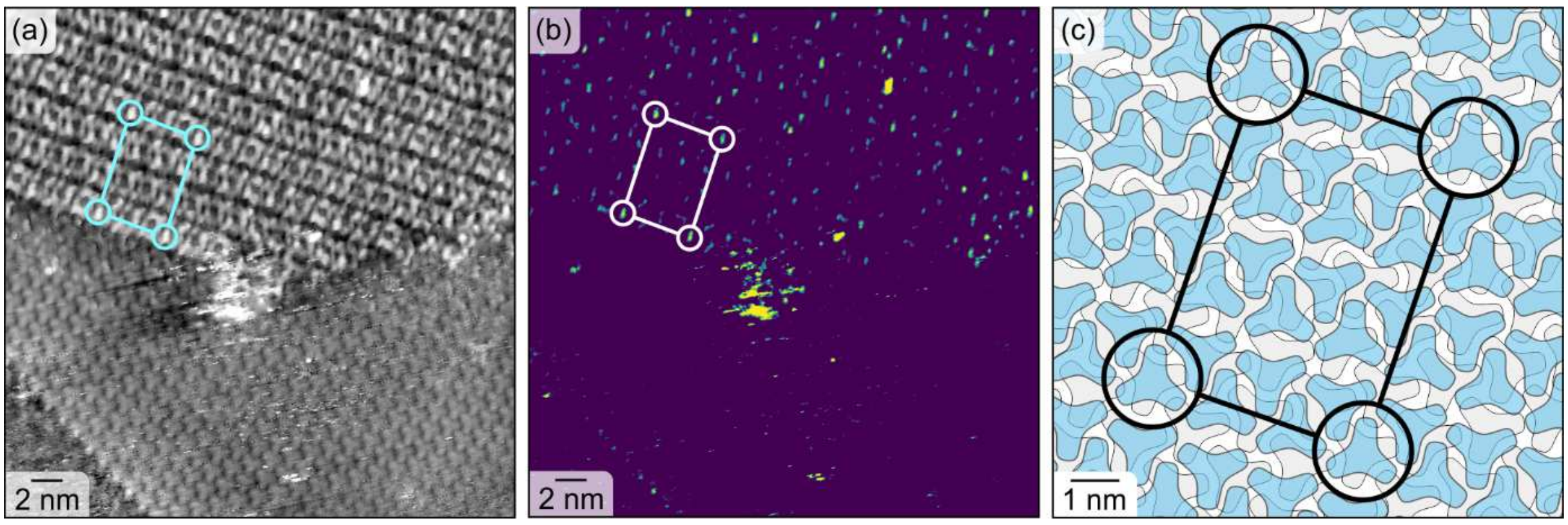

**Figure S8.** Periodicity of the BTB stripe phase on the δ-layer. (a) STM image of the BTB stripe phase on the δ-layer. Despite corrugation of BTB due to its structural adaptation to the substrate, the periodic motifs are clearly recognizable. (b) The same image with enhanced contrast and a marked unit cell. (c) Schematics of the BTB stripe phase on the δ-layer showing the mutual position of the BTB stripe overlayer (blue) and the underlying δ-layer (gray). Scanning parameters: 0.48 V, 50 pA.

## 9. Fitting parameters for O 1s peak

**Table S4**: Peak-fit parameters (XPS): binding energy position (BE) and FWHM of the Gaussian part of the O 1s peak components. The width of the Lorentzian component was fixed at 0.2 eV. The BE shifts (from the peak, marked +0.0) are given in brackets.

| | Component O1 (hydroxyl oxygen) | | Component O2 (carbonyl oxygen) | | Component O3 (carboxylate oxygen) | |
|---|---|---|---|---|---|---|
| | BE (eV) | FWHM (eV) | BE (eV) | FWHM (eV) | BE (eV) | FWHM (eV) |
| α | 533.75 (+1.6) | 1.5 | 532.15 (+0.0) | 1.5 | – | – |
| δ | – | | – | | 530.35 | 1.0 |
| α / δ | 533.40 (+1.6) | 1.5 | 531.80 (+0.0) | 1.3 | 530.35 | 1.0 |

The position of the Ag $3d_{5/2}$ peak was (368.24 ± 0.02) eV for all measurements.

## 10. Compressed semi-deprotonated stripe phase on Ag(111)

Upon annealing to temperatures below those required for complete deprotonation, a compressed semideprotonated stripe phase was observed on Ag(111). The extended unit cell is $\begin{pmatrix} 37 & 19 \\ 5 & 23 \end{pmatrix}$ with respect to the Ag(111) substrate; the size of the apparent unit cell is 20.18 × 15.44 Å with an angle of 77.18°. Corresponding structural and diffraction data are summarized in Figure S9.

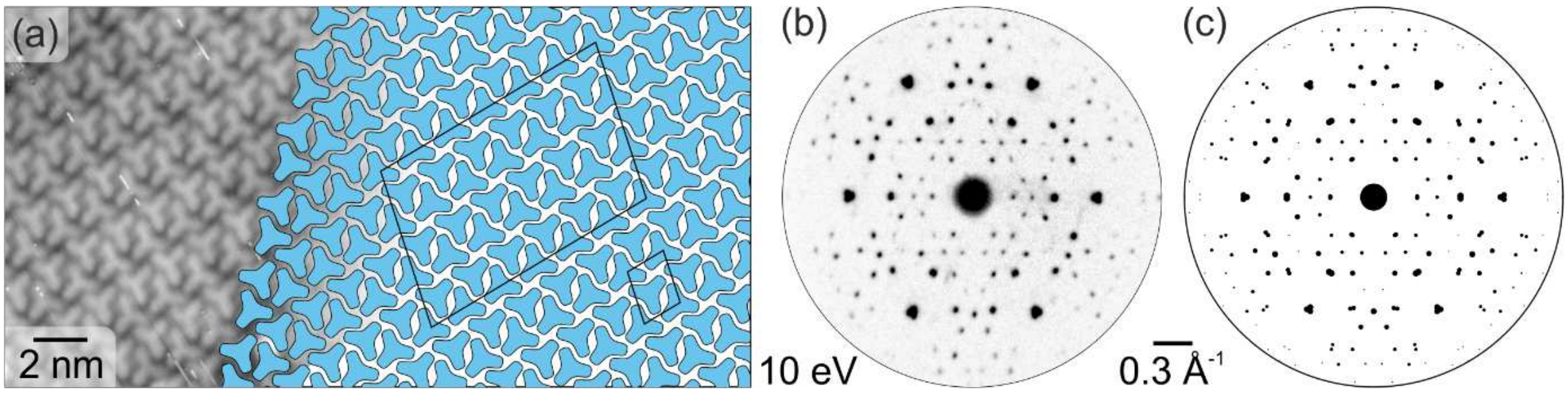

**Figure S9.** The semideprotonated γ-stripe phase on Ag(111). (a) STM image blended with schematics of the phase; parallelograms mark extended and apparent unit cells, respectively. Scanning parameters: 1.4 V, 50 pA. (b) Diffraction and (c) its model.

## 11. Peak positions with respect to the sample work function

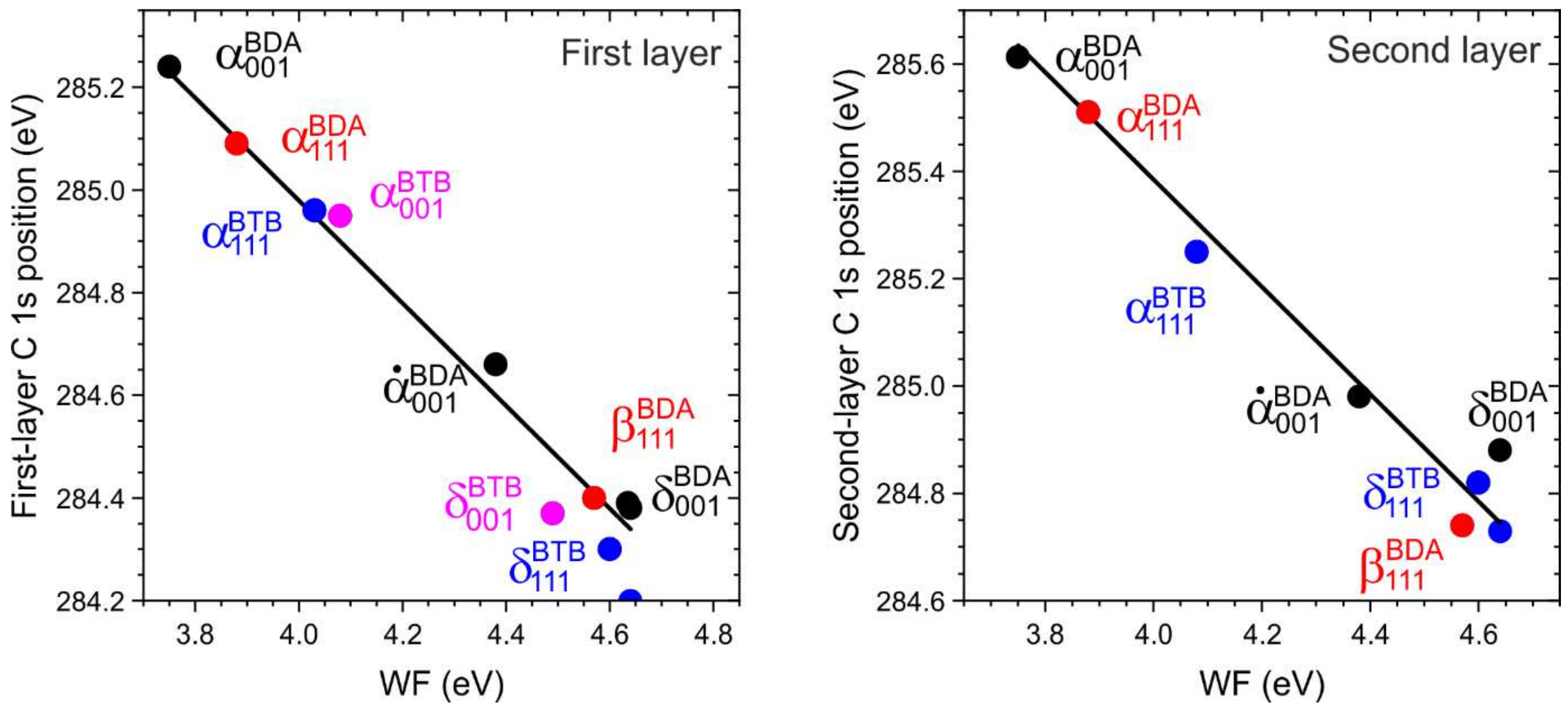

**Figure S10.** Position of the C 1s peak associated with phenyl rings within the first and second BTB molecular layers plotted as a function of the sample work function (WF) compared with earlier results for BDA.[18] The upper index marks the molecule, and the bottom index the Ag surface orientation. The lines have a slope of $(-1)$, the fitted experimental values have slope $-1.03 \pm 0.06$ (first layer) and $-0.93 \pm 0.08$ (second layer), respectively.

## 12. The models of unsupported honeycomb phase

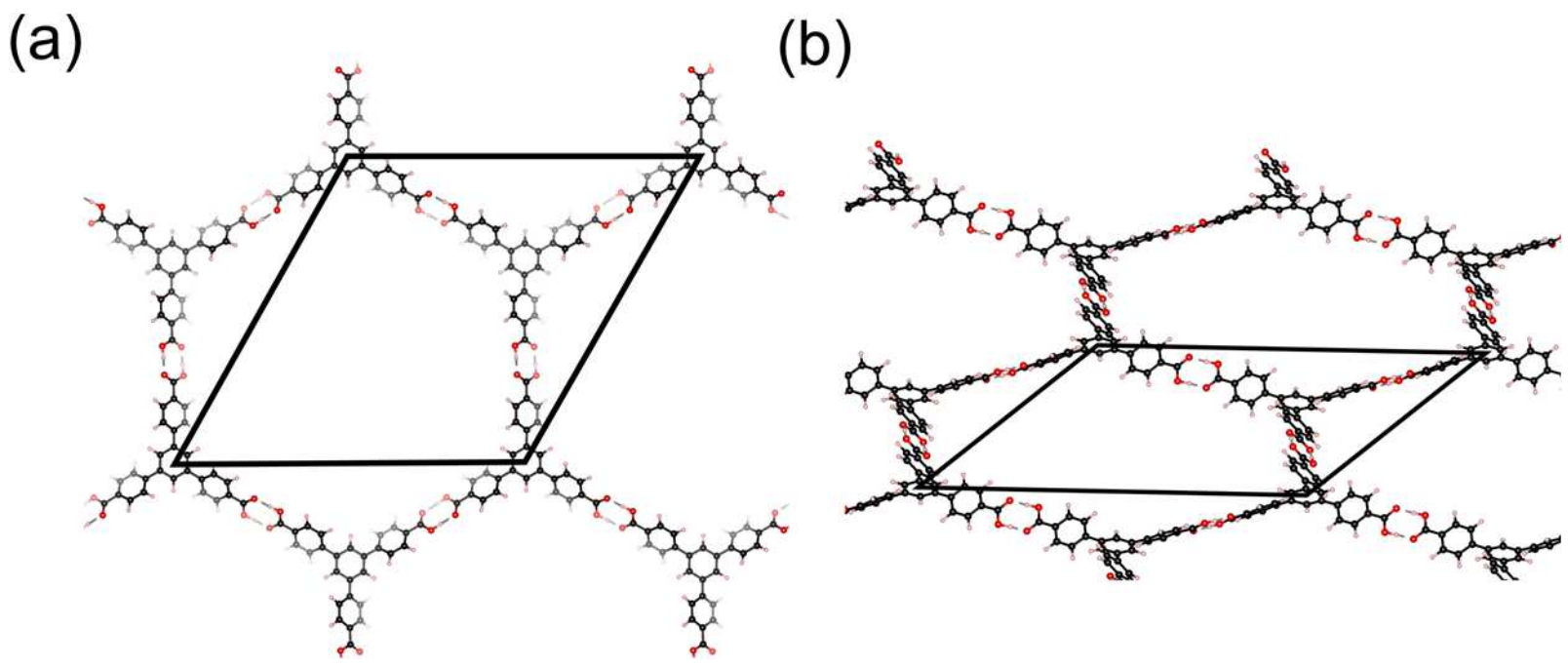

**Figure S11.** (a) top and (b) perspective view on the unsupported, corrugated honeycomb phase calculated with DFT.

## 13. Analysis of the unfavorable lateral positions for the BTB overlayer

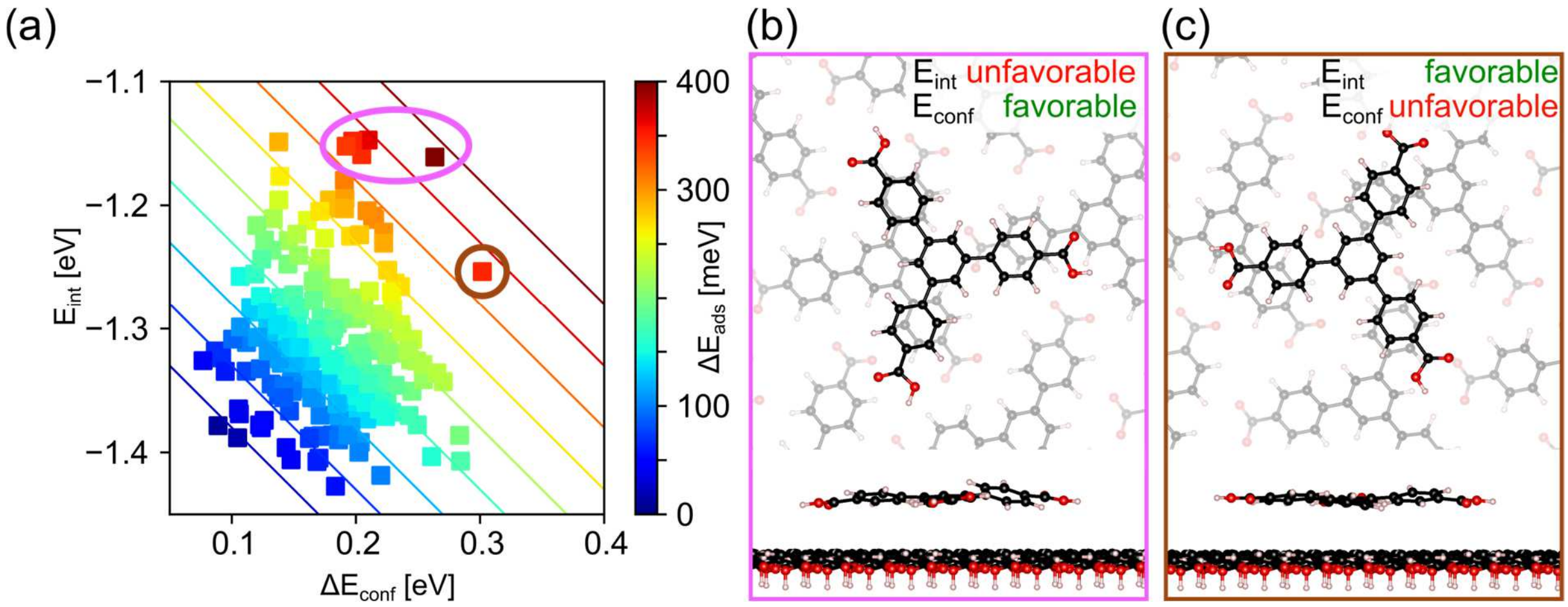

**Figure S12.** Extension of Figure 7 by the least stable geometries. Panel (a) shows relative values of $E_{\mathrm{int}}$ plotted against $E_{\mathrm{conf}}$ for all 240 individual BTB configurations, with the points colored according to relative adsorption energy $\Delta E_{\mathrm{ads}}$. The diagonal lines correspond to constant $E_{\mathrm{ads}}$ and show that a less favorable $E_{\mathrm{conf}}$ can be compensated by more favorable $E_{\mathrm{int}}$, and vice versa. While all unstable geometries show very similar lateral position on the δ-layer, with the edge phenyl rings directly overlapping with the phenyl rings in the substrate layer, molecular conformation changes from partially planarized geometries (b) to almost fully planarized geometries (c).

## 14. Suppression of the compensation mechanism in the optimized gas-phase BTB network

To assess the role of conformational adaptation in stabilizing the BTB heteroepitaxial structure, we performed an additional set of calculations in which the optimized gas-phase BTB network was placed on the δ-layer and subsequently relaxed. In this setup, the intermolecular interactions largely preserve the mutual positions of the BTB molecules, while their conformations can adapt only locally to the underlying layer. Comparison with the original dataset, shown in Figure S12, therefore allows us to quantify the energetic consequences of suppressing the full conformational adaptation to the δ-layer. For structures starting from the planar configuration (Figure S12a), the broad distribution of the conformation energy reflects the ability of individual molecules to deform in response to their local adsorption environment and compensate for differences in interaction energy. In contrast, structures starting from the optimized gas-phase geometry (Figure S12b) exhibit a substantially narrower range of $E_{\mathrm{conf}}$, shifted toward more favorable values. This indicates that the molecules remain closer to their preferred gas-phase conformations and undergo less pronounced conformational adaptation to the underlying δ-layer. Consequently, the molecules are less able to adapt to favorable adsorption geometries on the δ-layer, resulting in less favorable $E_{\mathrm{int}}$ and, on average, larger $\Delta E_{\mathrm{ads}}$.

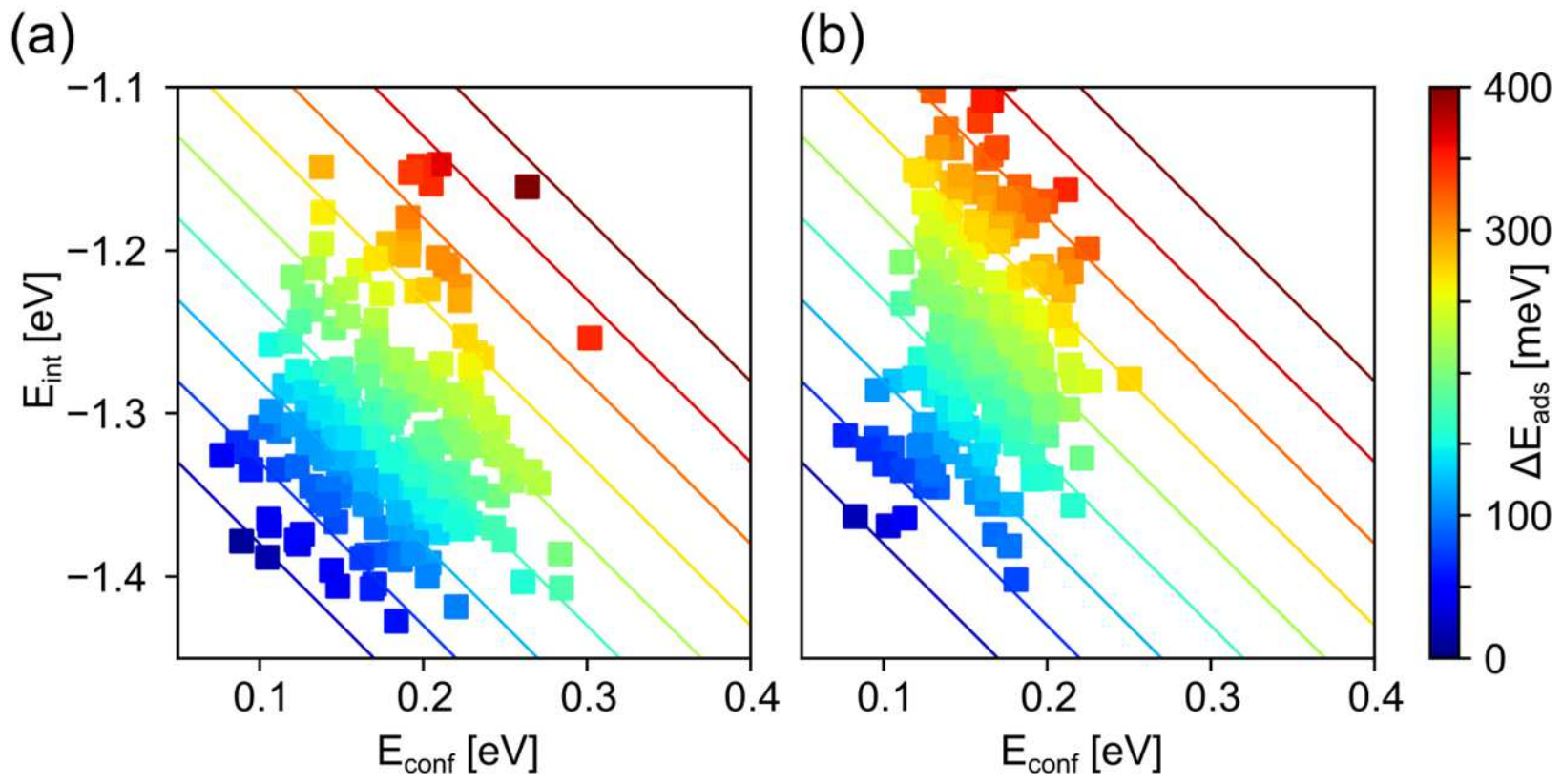

**Figure S13.** Effect of conformational adaptation on the balance between $E_{\mathrm{conf}}$ and $E_{\mathrm{int}}$. (a) Correlation between the conformational energy, $E_{\mathrm{conf}}$, and the interaction energy with the δ-layer, $E_{\mathrm{int}}$, for structures optimized starting from the planar configuration, corresponding to the dataset discussed in the main text. (b) The same analysis for structures optimized starting from the distorted gas-phase geometry. In (b), the considerably narrower distribution of $E_{\mathrm{conf}}$ demonstrates suppression of the compensation mechanism between molecular deformation and interaction with the underlying layer. Consequently, the structures cannot adapt efficiently to favorable adsorption geometries and exhibit, on average, less favorable $E_{\mathrm{int}}$ and higher $\Delta E_{\mathrm{ads}}$. The color scale represents the adsorption-energy difference, $\Delta E_{\mathrm{ads}}$, relative to the most stable configuration.

**15. Calculated Potential Energy Surface (PES) of the HMTP overlayer on the δ-layer**

Since the lattice parameters of the HMTP monolayer closely match those of the δ-layer (13.12 Å for HMTP and 13.23 Å for the δ-layer, corresponding to a lattice mismatch below 1%), the HMTP overlayer was modeled using a supercell containing one HMTP molecule and one δ-BTB molecule. Two orientational variants of the HMTP layer were considered. Both have the same unit cell and lattice periodicity but differ in the azimuthal orientation of the HMTP molecules, which are rotated by 30° with respect to the underlying δ-layer ($\gamma = 0°$ and 30° shown in Figure S14).

For PBE-D3, the potential energy surface was constructed from 900 structures sampling different lateral positions of the HMTP layer for each molecular orientation. To assess the sensitivity of the resulting PES to the treatment of exchange and correlation, an additional, sparser dataset of approximately 250 structures was calculated using the nonlocal $R^2$SCAN+rVV functional. The resulting PES maps for both functionals and both HMTP orientations are shown in Figure S14.

For both functionals and molecular orientations, the most favorable adsorption geometries are found when HMTP is located predominantly above the empty regions of the δ-layer. This behavior is consistent with that observed for the BTB overlayer. However, the detailed energetics depend on both the molecular orientation and the employed functional. The minimum of the $\gamma = 30°$ orientation (Figure S14c-d) is approximately 40 meV per HMTP molecule higher in energy than that of the $\gamma = 0°$ orientation (Figure S14a-b). This difference can be rationalized by the different registry of HMTP with the underlying δ-layer. For $\gamma = 0°$, most of the HMTP molecule can occupy the open region of the δ-layer while minimizing overlap with the underlying BTB molecules. In the $\gamma = 30°$ orientation, the molecular shape is less well aligned with this open region, resulting in greater spatial overlap with the δ-layer and a correspondingly less favorable interaction.

Comparison of the PBE-D3 and R²SCAN+rVV PES maps shows that the main features of the energy landscape are preserved: both functionals identify similar regions of particularly favorable and unfavorable adsorption. The overall corrugation of the PES, however, is functional dependent, increasing from approximately 200 meV with PBE-D3 to approximately 300 meV with R²SCAN+rVV. Consequently, some intermediate adsorption geometries appear comparatively favorable at the PBE-D3 level but become substantially less favorable with R²SCAN+rVV, as illustrated by the region highlighted by the circle in Figure S14a-b. These configurations involve partial overlap between aromatic rings of HMTP and the underlying BTB molecules. The difference therefore indicates that the relative energetics of such partially overlapping aromatic configurations are sensitive to the treatment of dispersion interactions, with the pairwise D3 correction and the nonlocal rVV treatment producing noticeably different energy landscapes.

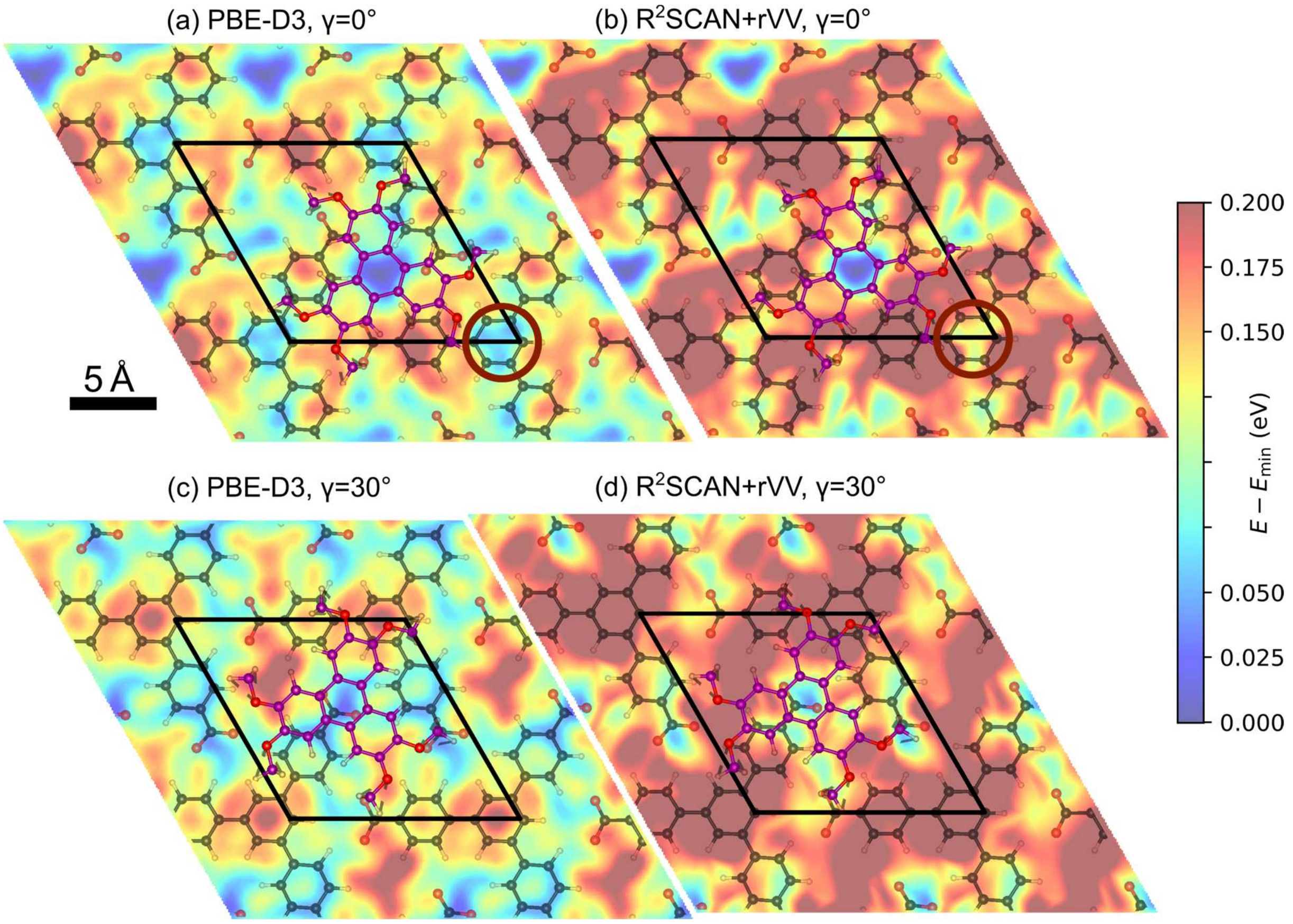

**Figure S14.** Calculated potential energy surfaces of the HMTP overlayer on the δ-layer for two

molecular orientations and two exchange-correlation treatments: (a) PBE-D3, $\gamma = 0°$; (b) $R^2$SCAN+rVV, $\gamma = 0°$; (c) PBE-D3, $\gamma = 30°$; and (d) $R^2$SCAN+rVV, $\gamma = 30°$. The two HMTP configurations have the same unit cell and lattice periodicity and differ only in the azimuthal orientation of the HMTP molecules, which are rotated by 30° relative to each other. The color scale shows the energy relative to the minimum-energy configuration for the corresponding PES, $E - E_{min}$. The black parallelogram indicates the unit cell. The circles in (a) and (b) highlight an adsorption region whose relative stability differs noticeably between PBE-D3 and $R^2$SCAN+rVV.

## 16. HMTP growth on Ag(111) and δ-layer

The structure of the HMTP (Figure S15a) on Ag(111) (Figure S15b) and the δ-layer (Figure S15c) is very similar: HMTP forms close-packed structure with a hexagonal lattice. The diffraction patterns of the δ-layer and HMTP/Ag(111), in Figures S15d and e are strikingly similar suggesting that both molecules have very similar superstructure when deposited on the Ag(111) surface; indicated by the same superlattice matrices and the unit cell vectors for both structures. In the case of MHTP/δ-layer, we do not observe any additional spots in the diffraction pattern in Figure S15f suggesting perfect epitaxial matching of HMTP on the δ-layer. Dark-field LEEM (Figure S15h) shows a higher intensity at the positions of HMTP islands (Figure S15g): a higher intensity is expected when two molecular layers (HMTP and δ-layer in this case) contribute to the intensity of the same diffraction spot.

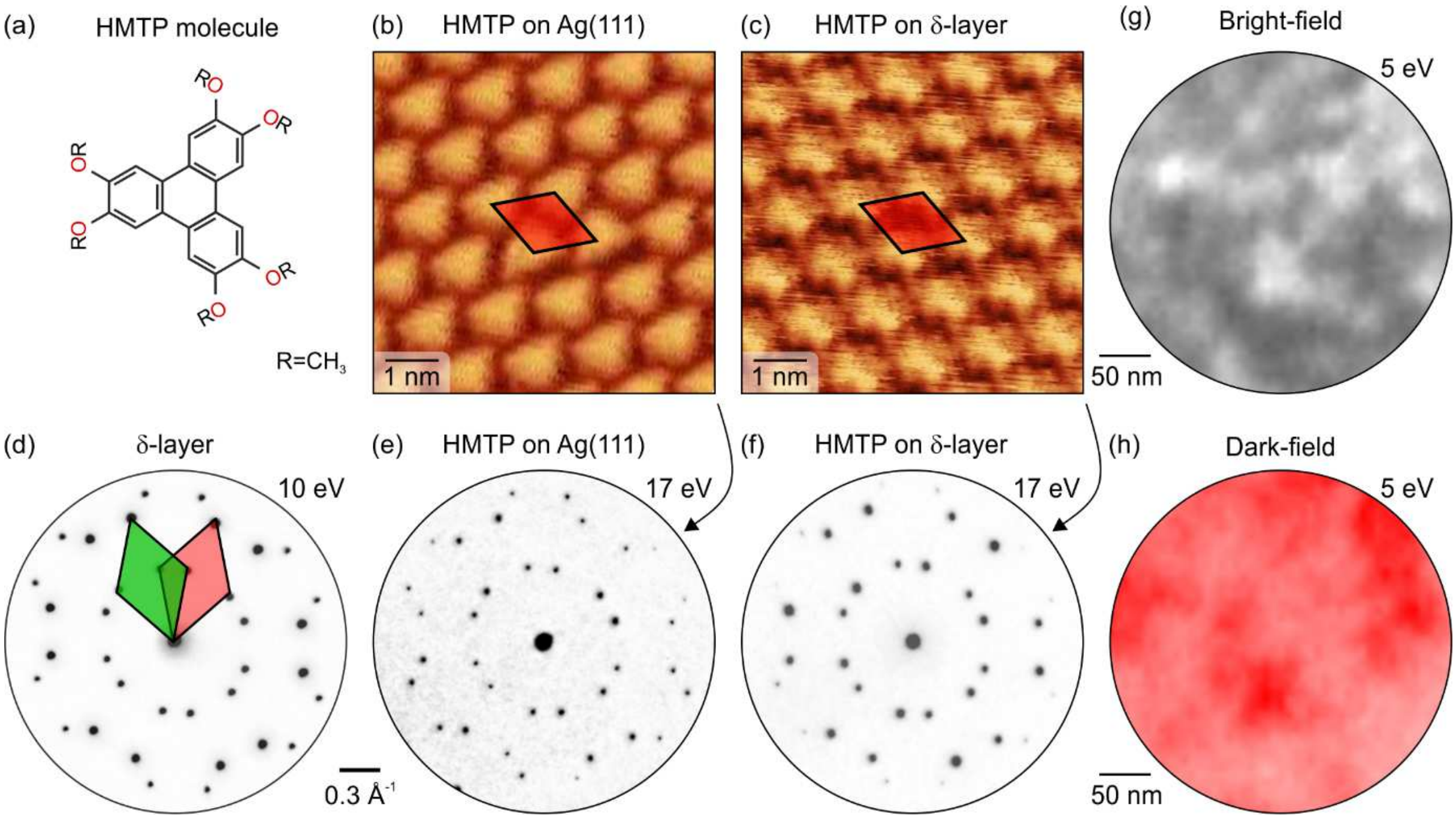

**Figure S15.** Analysis of HMTP on Ag(111) and δ-layer. (a) Molecular structure of HMTP. (b, c) STM image of monolayer HMTP on (b) Ag(111) and (c) on δ-layer. Note that as the HMTP has same unit cell as δ-layer, there is a minor chance that this is the δ-layer imaged under very specific tip conditions. (d, e, f) Diffraction measured at (d) δ-layer, (e) HMTP/Ag(111), and (f) HMTP/δ-layer. The parallelograms in (d) marks unit cells of two symmetry-equivalent

domains. (g) Bright and (h) dark-filed LEEM image showing HMTP islands on the δ-layer and their increased intensity in the dark-field.

## 17. Validity of the DFT models without the silver substrate

DFT calculations of the BTB overlayer were performed by removing the Ag substrate completely and saturating the underlying δ-layer with hydrogen atoms. To test the validity of this simplification, we calculated the isolated BTB molecule on top of the 2×2 superstructure of the δ-layer, laterally placed at distinct positions. The resulting benchmark data reveal that removing the silver substrate introduces error smaller than 15 meV, as shown in Figure S16.

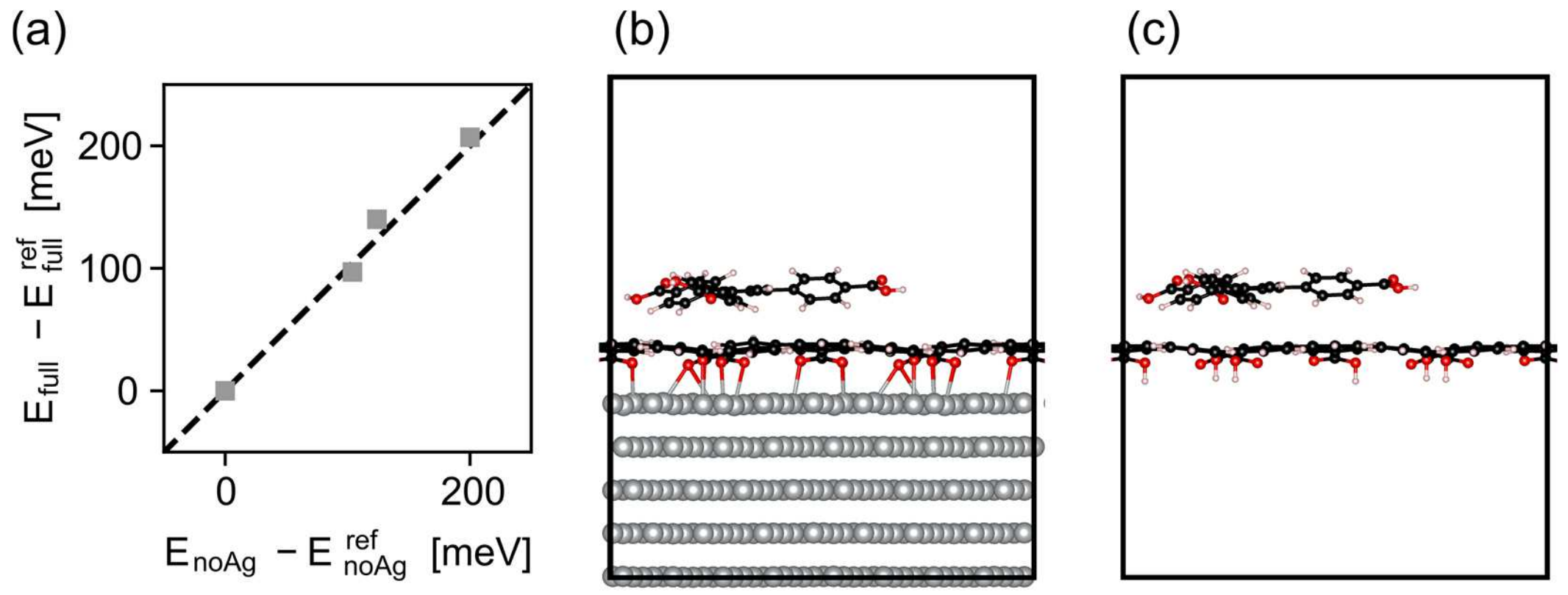

**Figure S16.** (a) Calculated total energies, referenced to the first structure of the dataset, in systems with and without the silver substrate. Model systems with and without the silver substrate used for benchmark calculations are further shown in (b) and (c), respectively.